\documentclass[useAMS,usenatbib]{mn2e}

\usepackage{psfig,graphicx}

\input epsf
\input psfig.sty
\usepackage{amsmath}
\usepackage{amssymb}
\usepackage{upgreek}
\usepackage{graphicx}

\newcommand{\arcs}{\mbox{\ensuremath{^{\prime\prime}}}}

\title[Extremely Red Stellar Objects]
  {Extremely Red Stellar Objects revealed by IPHAS}
\author[N.J.~Wright et al.]
  {N.J.~Wright,$^{1,2}$ R.~Greimel,$^{3,4}$ M.J.~Barlow,$^1$ J.E.~Drew,$^{5,6}$ M.-R.L.~Cioni,$^{6}$ \newauthor A.A.~Zijlstra,$^{7}$ R.L.M.~Corradi,$^{3}$ E.A. Gonz\'alez-Solares,$^8$ P.~Groot,$^{9}$ J.~Irwin,$^{2,8}$ \newauthor M.J.~Irwin,$^{8}$ A.~Mampaso,$^{10}$ R.A.H.~Morris,$^{11}$ D.~Steeghs,$^{2,12}$ Y.C.~Unruh,$^{5}$ N.~Walton$^{8}$  \\
  \\
  $^1$Department of Physics and Astronomy, University College London, Gower Street, London WC1E 
6BT, U.K.\\
  $^2$Harvard-Smithsonian Center for Astrophysics, 60 Garden Street, Cambridge, MA~02138, U.S.A.\\
  $^3$Isaac Newton Group of Telescopes, Apartado de correos 321, E-38700, Santa Cruz de la Palma, 
Tenerife, Spain\\
  $^4$Institut f\"ur Physik, Karl-Franzen Universit\"at Graz, Universit\"atsplatz 5, 8010 Graz, Austria\\
  $^5$Imperial College of Science, Technology and Medicine, Blackett Laboratory, Exhibition Road, 
London, SW7 2AZ, U.K.\\
  $^6$Centre for Astrophysics Research, University of Hertfordshire, College Lane, Hatfield, AL10 9AB, 
U.K.\\
  $^7$School of Physics and Astronomy, University of Manchester, Sackville Street, PO~Box~88, Manchester, M60~1QD, U.K.\\
  $^8$Institute of Astronomy, Madingley Road, Cambridge, CB3 0HA, U.K.\\
  $^9$Afdeling Sterrenkunde, Radboud Universiteit Nijmegen, Faculteit NWI, Postbus~9010, 6500~GL Nijmegen, the Netherlands\\
  $^{10}$Instituto de Astrofisica de Canarias, 38200 La Laguna, Tenerife, Spain\\
  $^{11}$Astrophysics Group, Department of Physics, Bristol University, Tyndall Avenue, Bristol, BS8~1TL, U.K.\\
  $^{12}$Department of Physics, University of Warwick, Coventry, CV4~7AL, U.K.\\
  }

\pagerange{\pageref{firstpage}--\pageref{lastpage}} \pubyear{2008}

\def\LaTeX{L\kern-.36em\raise.3ex\hbox{a}\kern-.15em
    T\kern-.1667em\lower.7ex\hbox{E}\kern-.125emX}

\begin{document}

\label{firstpage}

\maketitle

\begin{abstract}

We present photometric analysis and follow-up spectroscopy for a population of extremely red stellar objects extracted from the point-source catalogue of the INT Photometric H$\alpha$ Survey (IPHAS) of the northern galactic plane. The vast majority of these objects have no previous identification. Analysis of optical, near- and mid-infrared photometry reveals that they are mostly highly-reddened asymptotic giant branch stars, with significant levels of circumstellar material. We show that the distribution of these objects traces galactic extinction, their highly reddened colours being a product of both interstellar and circumstellar reddening. This is the first time that such a large sample of evolved low-mass stars has been detected in the visual and allows optical counterparts to be associated with sources from recent infrared surveys.

Follow-up spectroscopy on some of the most interesting objects in the sample has found significant numbers of S-type stars which can be clearly separated from oxygen-rich objects in the IPHAS colour-colour diagram. We show that this is due to the positions of different molecular bands relative to the narrow-band H$\alpha$ filter used for IPHAS observations. The IPHAS $(r' - $H$\alpha$) colour offers a valuable diagnostic for identifying S-type stars. A selection method for identifying S-type stars in the galactic plane is briefly discussed and we estimate that over a thousand new objects of this type may be discovered, potentially doubling the number of known objects in this short but important evolutionary phase.

\end{abstract}


\section{Introduction}

Evolved low and intermediate mass stars ($0.8 < M_{\odot} < 8.0$) in the asymptotic giant branch (AGB) stage of evolution are some of the most luminous stars in the Galaxy and represent one of the final evolutionary stages of all low and intermediate mass stars. Observations of these evolved stars show various degrees of photospheric abundance enhancements due to carbon and s-process elements being brought to the surface by the third dredge-up \citep{iben83}. In particular this has led to two specific subtypes of AGB stars known as S-type stars and carbon stars. S-type stars are thought to be the transition objects between the oxygen-rich M giants (C/O$ < 1$) and the carbon stars (C/O$ > 1$) as carbon is dredged up to the surface \citep{keen54}. The relative rarity of S-type stars has been attributed to the short transition time from oxygen-rich to carbon-rich surface chemistry, though there is evidence that this may not be the complete picture \citep[e.g.][]{chan91}.

Mass-loss during the AGB phase can replenish the interstellar medium (ISM) with this processed material and rates of mass-loss are observed to reach up to 10$^{-5}$~M$_{\odot}$yr$^{-1}$ at the tip of the AGB \citep[e.g.][]{knap85,zijl92}. The cool extended atmospheres of AGB stars are also one of the principle sites of dust formation in the Galaxy. The circumstellar material around such objects contributes to their reddening and, by the time they reach the tip of the AGB, they may become completely obscured at optical wavelengths. These stars are responsible for much of the processed material returned to the ISM and in doing so drive the evolution of the Galaxy \citep{buss99}.

The INT Photometric H$\alpha$ Survey of the northern galactic plane \citep[IPHAS,][]{drew05} is 
an imaging survey using the Wide Field Camera (WFC) on the Isaac Newton Telescope (INT). The imaging in broad-band Sloan $r'$ and $i'$ filters, accompanying narrow-band H$\alpha$ filter ($\lambda_c = 6568$~\AA, FWHM~=~95~\AA) observations, reaches to at least $r' = 20$. One of the survey's main aims is to identify large numbers of objects in important short-lived phases of stellar evolution \citep[see e.g.][]{with08,corr08}. The survey covers the entire northern galactic plane in the Galactic latitude range $-5^{\circ} \le b \le +5^{\circ}$, a total sky area of 1800 square degrees. More than 90\% of the survey area has already been observed and an initial release of data has been made \citep[see][]{gonz08}. The filters employed by IPHAS naturally allow it to highlight red and reddened populations in the galactic plane and it has uncovered large numbers of extremely red stellar objects (ERSOs), with ($r' - i'$) colour extending to almost $\sim$6, the majority of which have no previous identification. This work comprises a study of these objects to determine their typical properties and assess the cause of their highly reddened colours.

In Section~2 we present the IPHAS observations and select a population of extremely reddened stellar objects from the point source catalogue (PSC). In Section~3 we analyse the photometric properties of these sources to determine their typical properties and in Section~4 we discuss the small fraction of objects with previous identifications. In Section~5 we investigate the galactic distribution of the ERSO sample and compare with galactic reddening maps. In Section~6 we cross-correlate the IPHAS photometry of these objects with photometry from several recent infrared surveys, and assess the fraction of objects with infrared excesses. In Section 7 we present near-IR spectroscopy of some of the most extreme sources in the colour-colour diagram, including the discovery of several S-type stars in a distinctive region of the IPHAS colour-colour plane. Finally, in Section 8 we outline a method for separating these interesting objects from the complete IPHAS catalogue.

\section{Selection of extremely red stellar objects from the IPHAS catalogue}
\label{s-ersos}

\begin{figure}
\begin{center}
\includegraphics[width=175pt, angle=270]{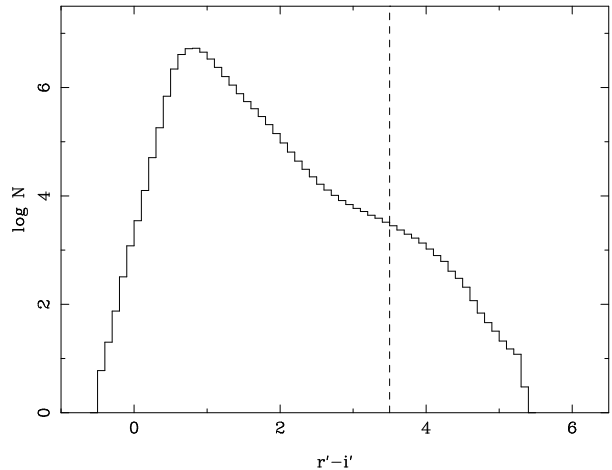}
\caption{The ($r' - i'$) distribution of all objects in the IPHAS catalogue that fulfill the selection criteria listed in the text, binned into 0.1~magnitude bins. The arbitrary selection level defining extremely red stellar objects (ERSOs), ($r' - i') > 3.5$, is indicated with a dashed line.}
\label{ri_magnitudes}
\end{center}
\end{figure}

\citet{drew05} showed that the IPHAS $(r' - $H$\alpha, r' - i')$ colour-colour diagram may be used to pick out many different types of objects and that the predicted spectral-type trends are congruent with initial results of follow-up spectroscopy. The main-sequence and giant branch can be separated in the colour-colour plane for stars cooler than $\sim$5000~K, potentially allowing the galactic population of cool giants to be isolated. However, the effects of circumstellar and interstellar reddening play an important role in determining the position of stars in the colour-colour diagram and the IPHAS catalogue contains a large number of sources with colours up to and in excess of ($r' - i') \sim 5$ (see Figure~\ref{ri_magnitudes}).

We have embarked upon a photometric and spectroscopic study of the extreme IPHAS objects with $(r' - i') > 3.5$ (an arbitrary cut-off chosen to give a sizable working catalogue; see Figure~\ref{ri_magnitudes}). This group of ERSOs was compiled from all IPHAS observations taken between August~2003 and January~2007. The quality criteria used for these data were: seeing less than 2~arcsec; ellipticity less than 0.2; and $r'$-band sky counts less than 2000~ADUs \citep{gonz08}. From these we applied the following further quality criteria to all sources:

\begin{enumerate}
\item they must have been detected at least twice in the survey (usually in a direct field and its offset field);
\item detections must be classified as "stellar" or "probably stellar" \citep[c.f.][]{gonz08} in the $r'$, $i'$ and H$\alpha$ bands (for a small number of sources with "stellar" and "non-stellar" classifications on separate observations, visual inspection was used to re-classify the data as necessary);
\item the mean $(r' - i')$ colour from all valid detections must be greater than 3.5.
\end{enumerate}

Requirement (i) is intended to remove erroneous detections, while (ii) will remove galaxies and nebulae with reddened colours (since we only intend to examine the stellar objects). The need for offset fields to fill in sky areas missed by gaps between the CCDs means that the majority of objects ($\sim$95\%) will have been observed at least twice, with some re-observed fields leading to more than two observations. If repeat observations are separated by more than a day, the potential for variability to influence magnitudes is increased. Our quality criteria ignore this, since secular magnitude changes do not necessarily significantly affect colours, and indeed variability of this nature may aid source classification.

\section{Colour-colour and colour-magnitude diagrams of the ERSOs}
\label{s-colours}

In this section we use IPHAS photometry to analyse the general properties of the ERSO sample and determine the range of spectral types that may contribute to the sample and the amount of reddening they may require to be included in the sample.

\begin{figure}
\begin{center}
\includegraphics[width=175pt, angle=270]{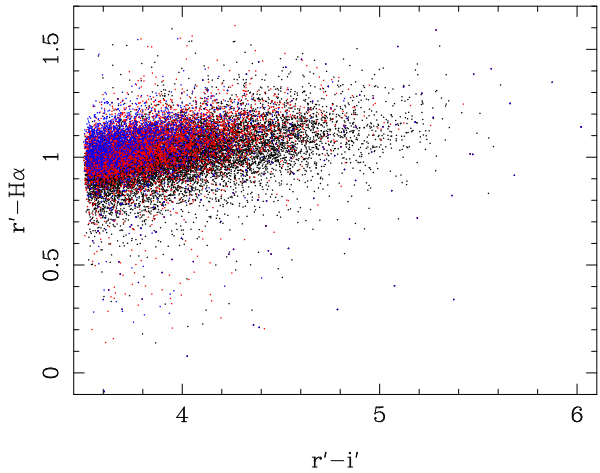}
\caption{The extremely red region of the IPHAS colour-colour plane, showing the 25,473 objects with (r
$'$~-~$i'$)~$> 3.5$. Colours indicate $r'$-band magnitudes split into three groups: $14 < r' \leq 17$ (blue), $17 < r' \leq 18.5$ (red) and $r' > 18.5$ (black).}
\label{ERSO_colcol}
\end{center}
\end{figure}

\begin{figure}
\begin{center}
\includegraphics[width=175pt, angle=270]{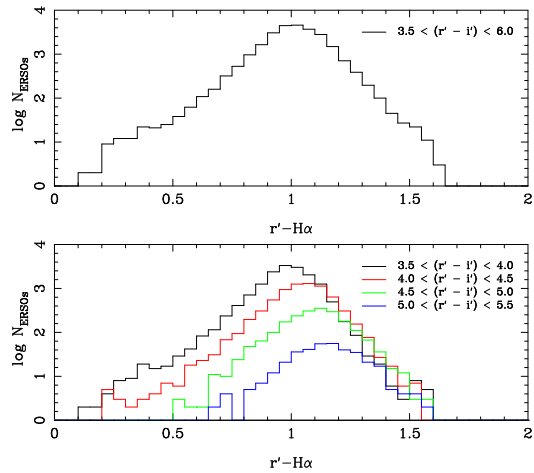}
\caption{Histogram of the number of ERSOs in ($r' - $H$\alpha$) bins of 0.05 magnitudes. {\it Top panel:} All ERSOs. {\it Bottom panel:} All the ERSOs separated into four groups as per their ($r' - i'$) colour. The vertical scales in both plots are logarithmic.}
\label{rha_mags}
\end{center}
\end{figure}

The positions of all 25,473 ERSOs in the IPHAS $(r' - $H$\alpha, r' - i')$ colour-colour plane can be seen in Figure~\ref{ERSO_colcol}. These objects are predominantly clustered along a narrow, slightly inclined band at ($r'$-H$\alpha) \sim 1$. However, there is a considerable width to this band, the sample showing variation from below zero to greater than 1.5 (see Figure~\ref{rha_mags}). A part of this spread is probably due to photometric errors in the H$\alpha$ band at the faint end of the IPHAS PSC, though it is unlikely to be responsible for the full range of $(r' - $H$\alpha$) values seen.

The total histogram of ($r' - $H$\alpha$) colours is shown in the upper panel of Figure~\ref{rha_mags} where it can be seen to be skewed, peaking at ($r' - $H$\alpha) \simeq 1$. The lower panel shows the spread in ($r' - $H$\alpha$) in four separate groups of ($r' - i'$) colour: all four contain hints of a second second, smaller peak at a lower ($r' - $H$\alpha$) colour. We will suggest in Section~\ref{s-extract} that this may be due to the galactic population of S-type stars.

In the main, the ERSOs are likely a combination of intrinsically red objects and highly reddened objects of all spectral types. The intrinsically red objects include late-type dwarf stars, red giant branch (RGB) and asymptotic giant branch (AGB) stars and potentially some red supergiants. We can immediately exclude significant numbers of late-type main sequence stars in the ERSO sample on account of their intrinsic colours and faint absolute magnitudes. Essentially, M dwarfs are so faint that IPHAS imaging only picks them up to distances under 100~pc. And in order for $(r' - i')$ to exceed 3.5, these objects would have to suffer at least $\sim~8$ magnitudes of extinction (for M6V dwarfs). Together these constraints imply an absurdly high differential extinction ($A_V \geq 100$ per kpc, as a rough lower bound).

\begin{figure*}
\begin{center}
\includegraphics[width=150pt, angle=270]{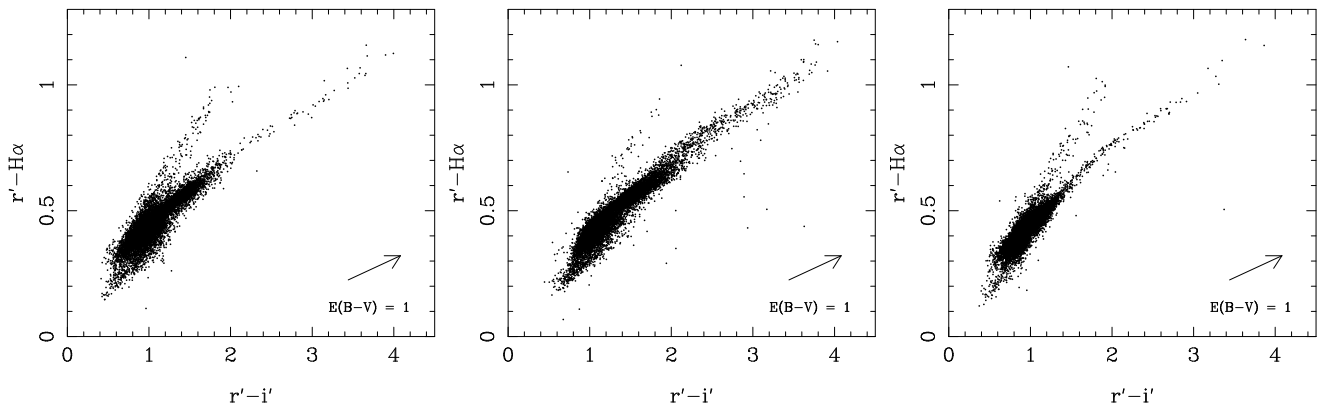}
\caption{IPHAS colour-colour diagrams for fields 5419 (in Cassiopeia), 4128 and 4150 (both in Scutum) showing all sources with $r' < 19$. The main sequence (upper track) and giant branch (lower, redder track) are clearly visible and separate at red colours. Those sources with ($r' - i') > 3.5$ are members of the ERSO sample. Approximate reddening vectors are shown for E(B-V)~$= 1$.}
\label{fields_cc}
\end{center}
\end{figure*}

Figure~\ref{fields_cc} shows colour-colour diagrams for three IPHAS fields with well-developed giant branches, each including objects that meet the criteria for ERSOs. The main sequence and giant branches extend to higher $(r' - $H$\alpha)$ with decreasing stellar temperature because the absorption from TiO in the $r'$ band increases \citep{drew05}. In stars of higher luminosity class the TiO bands are wider and cause absorption in the H$\alpha$ filter, which causes late-type stars of higher luminosity to have lower $(r' - $H$\alpha$) colour. The main-sequence clearly does not extend to sufficiently reddened colours for main-sequence stars to be classified as ERSOs. The two sequences separate around $(r' - i') \sim 1$, at approximate spectral type M0III on the giant branch. In the IPHAS colour-colour diagram the giant branch thereafter reddens onto itself \citep[c.f.][]{drew05}. Variations in the $(r' - i')$ colour along the giant branch are thus as much as indicator of temperature class as of reddening. Stars of higher luminosity classes are expected to be almost coincident with those of giant stars in the colour-colour diagram, but due to their relative scarcity are not expected to contribute significantly to the ERSO sample.

\begin{figure}
\begin{center}
\includegraphics[width=175pt, angle=270]{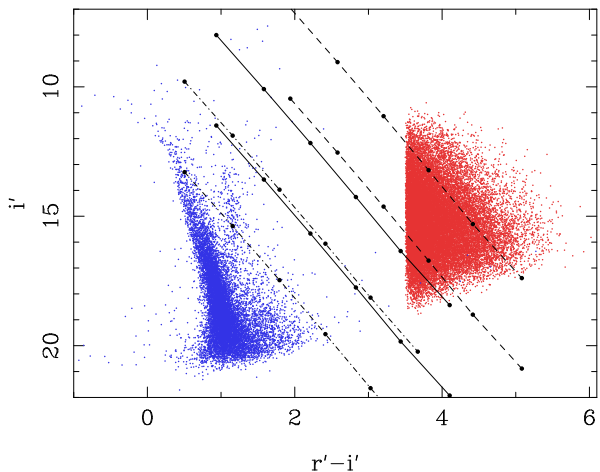}
\caption{IPHAS colour-magnitude diagram for all ERSOs (red points), compared to IPHAS sources in field 2963, in Cepheus (blue points). Overlaid are reddening loci for three types of giant star: K0 (dash-dotted line), M0 (full line) and M5 (dashed line), for distances of 1kpc (upper line) and 5kpc (lower line). Points along the reddening loci indicate unit intervals of E(B-V) from 0 to 5.}
\label{eros_cmd}
\end{center}
\end{figure}

The majority of ERSOs are therefore expected to be giant stars, either on the red giant branch or the asymptotic giant branch. Figure~\ref{eros_cmd} shows an $i'$ vs $(r' - i')$ colour-magnitude diagram for all the ERSOs, as well as all the sources from a typical IPHAS field. Overlaid are reddening loci for three types of giant star: K0, M0 and M5. Absolute visual magnitudes were taken from \citet{schm82}, colours from \citet{bess88} and converted onto the Wide Field Survey (WFS) Sloan system using the transformations given on the Cambridge Astronomical Survey Unit Wide Field Survey webpages\footnote{http://www.ast.cam.ac.uk/$\sim$wfcsur/technical/photom/colours/}. The limiting magnitudes of the survey in the $r'$ and $i'$ bands are evident in the positions of the faintest sources across the $(r' - i')$ colours.

The positions of the ERSOs relative to the reddening loci indicate that they are most probably M-type giant stars, which would be found on the asymptotic giant branch. Earlier spectral types, such as K-type giants, are unlikely to significantly contribute to the sample as they will be too faint and not sufficiently reddened to become classified as an ERSO. An M0 (M5) giant star will require a reddening of E(B-V)~$> 4$~(2) to become an ERSO (or $A_v > 12$~(6)~mags), while a reddening greater than E(B-V)~$\sim 5$~(6) will make the source too faint to be detected by IPHAS in the $r'$ band. This reddening may in general be due to a combination of interstellar dust, and newly created dust in the star's own ejecta.

\section{Previously identified objects}
\label{s-simbad}

\begin{figure}
\begin{center}
\includegraphics[width=175pt, angle=270]{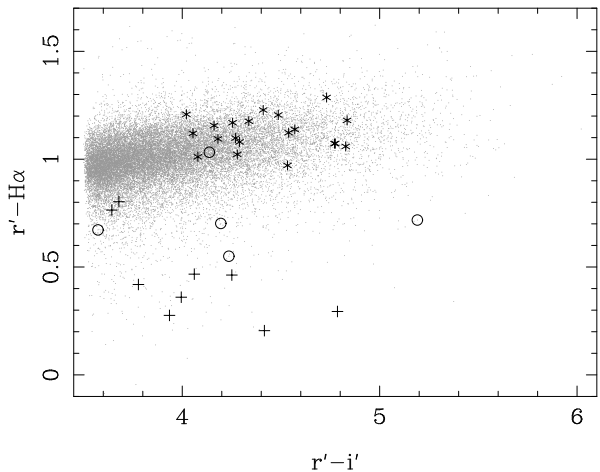}
\caption{The extremely red region of the IPHAS colour-colour plane with previously identified objects marked. These objects were identified from a SIMBAD search of the ERSO database using a radius of 3~arcsec. Carbon stars are shown as circles, S-type stars are plus symbols and OH/IR stars are star symbols.}
\label{simbad_search}
\end{center}
\end{figure}

An early spectroscopic analysis of stars near the ERSO region by \citet{drew05} revealed the presence of a mid-M giant and a carbon star. While these were not strictly in the ERSO region as we defined above (their ($r' - i'$) colours were 3.39 and 2.97 respectively), they provide a useful indication of the types of object prevalent in this area. We used the SIMBAD\footnote{http://simbad.u-strasbg.fr/simbad/} Astronomical Database to search for previously identified objects within our ERSO sample. Using a search radius of 3~arcsec we found identifications for 223 objects within our sample, less than 1\% of the 25,473 ERSOs. 187 of these were variable and included 50 of Mira type. Their distribution in $(r' - $H$\alpha)$ follows that of Figure~\ref{ERSO_colcol}. In addition there were 20 OH/IR stars, all with $(r' - i') > 4$ and confined to the densely populated $(r' - $H$\alpha$) band. Our search also uncovered 5 carbon stars and 9 S-type stars. We plot the positions of these last 3 object types in the colour-colour diagram, superposed on the ERSO population, in Figure~\ref{simbad_search}.

All nine known S-type stars - identified from objective prism surveys \citep{step90} - appear below the main locus, with all but two appearing significantly lower than the preponderance of sources. Four of the five carbon stars also appear below the main locus, though higher than the distribution of S-type stars. One carbon star lies higher, within the main locus. This is V2230~Cyg, a carbon star listed in the General Catalog of Galactic Carbon Stars \citep{alks01} that was identified as such by V.~Blanco (no separate publication). Its identification as a carbon star therefore cannot be properly verified. Its near-IR 2MASS photometry ($J-H = 1.2$, $H-K = 0.6$) places it in a region which includes both carbon stars and M-type supergiants \citep[e.g.][]{zijl96}, so its true identity remains unknown. The remaining carbon stars were either listed in the General Catalogue of Cool Carbon Stars \citep{step73} or classified on IRAS low resolution spectra by \citet{kwok97}.

The lower $(r' - $H$\alpha)$ values of the known carbon and S-type stars relative to M giants with more normal surface abundances is indicative of weak or absent TiO bands in the red spectrum (see Figure~19 of \citealt{drew05}; Section~\ref{s-extract} and Figure~\ref{bands_compare}). Our SIMBAD search also revealed one young stellar object (YSO) with $(r' - i') = 4.1$ and $(r' - $H$\alpha) = 0.8$ identified by \citet{fell00} based on ISOGAL mid-infrared photometry, as well as an M5 supergiant, VLH96~A \citep{vrba00}. The presence of these objects indicates that our sample may contain a small contamination of objects of earlier spectral type and higher luminosity class, but the considerable reddening required for the former and the scarcity of the latter are unlikely to make either a major contributor to our ERSO sample.

\section{The galactic distribution of highly-reddened sources}
\label{s-galactic}

\begin{figure}
\begin{center}
\includegraphics[height=230pt, angle=270]{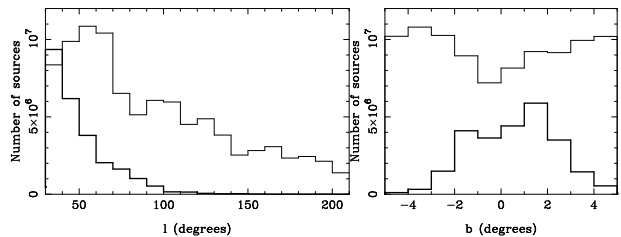}
\caption{Galactic distributions for all IPHAS sources (dashed line) and the ERSO sample (thick black line), divided by the number of IPHAS fields in each bin. {\it Left:} Distribution in galactic longitude for $30 < l < 210$ in $10^{\circ}$ bins. {\it Right:} Distribution in galactic latitude for $-5 < b < +5$ in $1^{\circ}$ bins. The number of ERSOs has been multiplied by 1000 on both plots to allow both sets of data to be visible.}
\label{histogram_galactic}
\end{center}
\end{figure}

Figure~\ref{histogram_galactic} shows the distribution of ERSOs in both galactic longitude and galactic latitude, compared to the distribution of all IPHAS sources. Unlike younger sources, old low- to intermediate-mass evolved stars will not trace galactic structure in the form of spiral arms and star-forming regions, but should show a distribution dependent only on galactocentric distance and disk scale height.

The longitude distribution of all IPHAS sources (left panel, Figure~\ref{histogram_galactic}) shows a peak at $l \sim 55^{\circ}$ as noted by \citet{with08}. This is thought to correspond to the position of the Sagittarius-Carina spiral arm \citep[e.g.][]{russ03}. From around $70^{\circ}$ the total number of sources begins to decline, with a dip around $l \sim 80^{\circ}$, probably associated with high extinction in the Cygnus region. The distribution of ERSOs shows a very different structure, with an exponential drop-off of sources towards increasing longitudes.

The latitude distribution of ERSOs is much more peaked towards the galactic equator than that of all IPHAS sources (right panel, Figure~\ref{histogram_galactic}), and peaks at a higher positive latitude. \citet{with08} noted similar properties in comparing the latitude distribution of IPHAS H$\alpha$-emitting sources with that of all IPHAS sources (their Figure~4) suggesting that the non-central latitude distribution peak was evidence for the Galactic warp \citep[e.g.][]{freu94}.

The concentration of ERSOs towards the galactic equator and towards lower longitudes in Figure~\ref{histogram_galactic} can be attributed to the denser sight-lines through the galactic disk in these directions. That this effect is not echoed in the distribution of all IPHAS sources is a result of the typical properties of the sources making up each sample. While the general IPHAS population is dominated by intrinsically faint G and K dwarfs, the ERSO and H$\alpha$-emitting samples are drawn from a fundamentally more luminous population visible over a larger range of distances in the galactic thin disk.

\begin{figure}
\begin{center}
\includegraphics[height=240pt, angle=270]{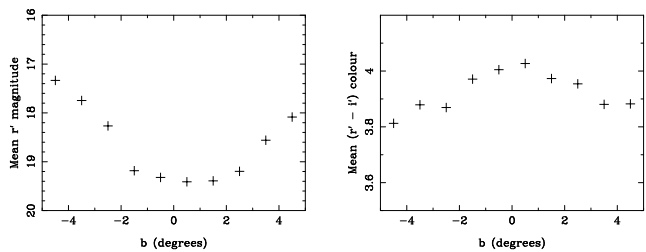}
\caption{Mean $r'$ magnitude (left) and ($r' - i'$) colour (right) from IPHAS photometry of the ERSO sample in 1$^{\circ}$ galactic latitude bins.}
\label{mean_magnitude}
\end{center}
\end{figure}

Both the ERSO and complete IPHAS samples show a small dip in the latitude bin from $-1$ to $0$, which we attribute to the extremely high extinction in the galactic mid-plane. This view is reinforced by the mean $r'$ magnitude of ERSOs shown in Figure~\ref{mean_magnitude} (left panel). The fainter mean magnitude and redder mean $(r' - i')$ colour (Figure~\ref{mean_magnitude}, right panel) of sources at low galactic latitudes support the interpretation that the ERSO distribution is magnitude-limited at low latitudes and volume-limited at higher latitudes.

\begin{figure*}
\begin{center}
\includegraphics[height=520pt, angle=270]{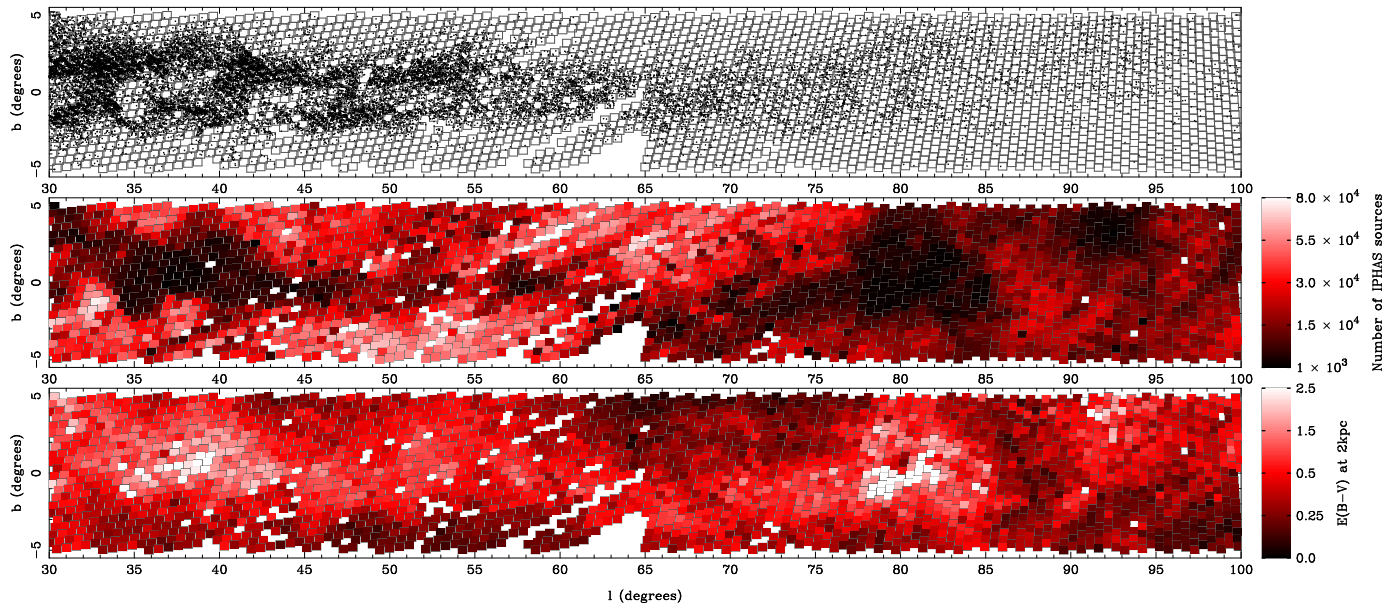}
\caption{The distribution of ERSOs in the galactic plane for $30 < l < 100$ compared to all IPHAS sources and compared to the distribution of interstellar extinction. The positions of the IPHAS fields are illustrated as grey rectangles with unobserved fields or fields that did not meet the quality criteria outlined in Section~\ref{s-ersos} left empty. {\it Top:} The positions of all ERSOs shown as black points. {\it Middle:} The total number of IPHAS sources in each IPHAS pointing box is illustrated via the colour of each box with unobserved or poor quality fields left white. {\it Bottom:} Galactic interstellar reddening out to a distance of 2~kpc, extracted from \citet{mars06}. Unobserved or poor-quality fields are coloured white.}
\label{inner_galactic}
\end{center}
\end{figure*}

To further investigate the distribution of ERSOs in the galactic plane, in Figure~\ref{inner_galactic} we have compared the distribution of all ERSOs in the galactic plane for $30 < l < 100$, to that of all IPHAS sources. Also shown is the galactic interstellar reddening, E(B-V) out to a distance of 2~kpc, extracted from the data presented by \citet{mars06}. A comparison between the prevalence of IPHAS sources and the interstellar extinction map shows a strong correlation between regions of high extinction and low IPHAS star counts, as would be expected. At a distance of 2~kpc some areas of the galactic plane reach reddening levels of E(B-V)~=~2.5, equivalent to $\sim$7~magnitudes of extinction in the $r'$~band, sufficient to obscure many main-sequence stars. The distribution of ERSOs shows the opposite trend however, with the extremely red objects confined to regions of the galactic plane with high extinction.

The correlation of ERSOs with interstellar extinction and the anti-correlation with IPHAS sources appears very strong and leads to the explanation that the extremely red colours of the majority of the ERSOs are due to interstellar reddening. As was shown in Section~\ref{s-colours}, late-type giant stars will require a reddening of E(B-V)~$\sim 2-6$ to become sufficiently reddened for inclusion in the ERSO sample. The areas of the galactic plane with E(B-V)~$> 2$ are almost completely coincident with those areas containing ERSOs. Some regions of the plane where galactic reddening is highest (e.g. around $l = 30-35^{\circ}$, $b = 0^{\circ}$) show lower numbers of sources than in neighbouring high reddening regions. The most plausible explanation for this is that the extinction is so high in these regions that even the deep IPHAS photometry cannot detect such reddened objects.

\section{Cross-correlation with existing infrared catalogues}
\label{s-infrared}

This section is a comparison of the ERSO sample with data from the large range of existing infrared catalogues to assess the incidence and properties of ERSO circumstellar emission. In particular the amount of circumstellar material typical of these can be parameterised in terms of an infrared colour excess derived from the available photometric databases.

\begin{table*}
\caption{Infrared catalogues cross-correlated with the 25,473 IPHAS ERSOs. Numbers of matches include all detections, at any of the photometric points, regardless of quality. The percentage of ERSOs with infrared detections (column 6) applies to all ERSOs, including those in regions not covered by all the infrared catalogues listed.}
\begin{tabular}{@{}lllcrc}
\hline
Catalogue & Photometric points & Sky coverage & Match radius & \multicolumn{2}{c}{Matches} \\
 & ($\mu$m) & & (arcsec) & Number & \% of ERSOs \\
\hline
2MASS & 		1.24 ($J$), 1.66 ($H$), 2.16 ($K_s$)	 &	All-sky		& 0.5 & 25411 & 99.8 \\
IRAS	 &		12, 25, 60, 100 &					All-sky		& 5 & 2884 & 11.3 \\
MSX & 		8.28, 12.13, 14.65, 21.34 &	$l = 0-360$, $-5 < b < +5$ 	& 2 & 9407 & 37.1 \\
ISOGAL & 	1.24 ($J$), 2.16 ($K$), 7, 15 &	Selected fields in the galactic-plane  & 6 & 102 & ~0.4 \\
GLIMPSE & 	3.6, 4.5, 5.8, 8.0 &		$l = 10-65$, $-1 < b < +1$	 & 1 & 4397 & 17.3 \\
\hline
\end{tabular}
\label{catalogues}
\end{table*}

We cross-correlated the ERSO sample of IPHAS sources with objects in the 2 Micron All Sky Survey 
\citep[2MASS,][]{skru06}, the Infrared Astronomical Satellite survey \citep[IRAS,][]{neug84}, the 
Midcourse Space Experiment survey \citep[MSX,][]{egan96}, the Infrared Space Observatory (ISO) 
survey of the inner galaxy \citep[ISOGAL,][]{omon03} and the Galactic Legacy Infrared Mid-Plane Survey Extraordinaire \citep[GLIMPSE,][]{benj03}.  Statistics for the cross-correlations are given in Table~\ref{catalogues}. The number of matches is dependent on the sky coverage of each survey, the photometric depth and, for mid- and far- infrared surveys, the number of ERSOs with circumstellar emission due to dust. Match radii, specified below, were chosen to be the largest value that avoided the frequent inclusion of more than one counterpart per ERSO.

\subsection{2MASS}

\begin{figure*}
\begin{center}
\includegraphics[width=280pt, angle=270]{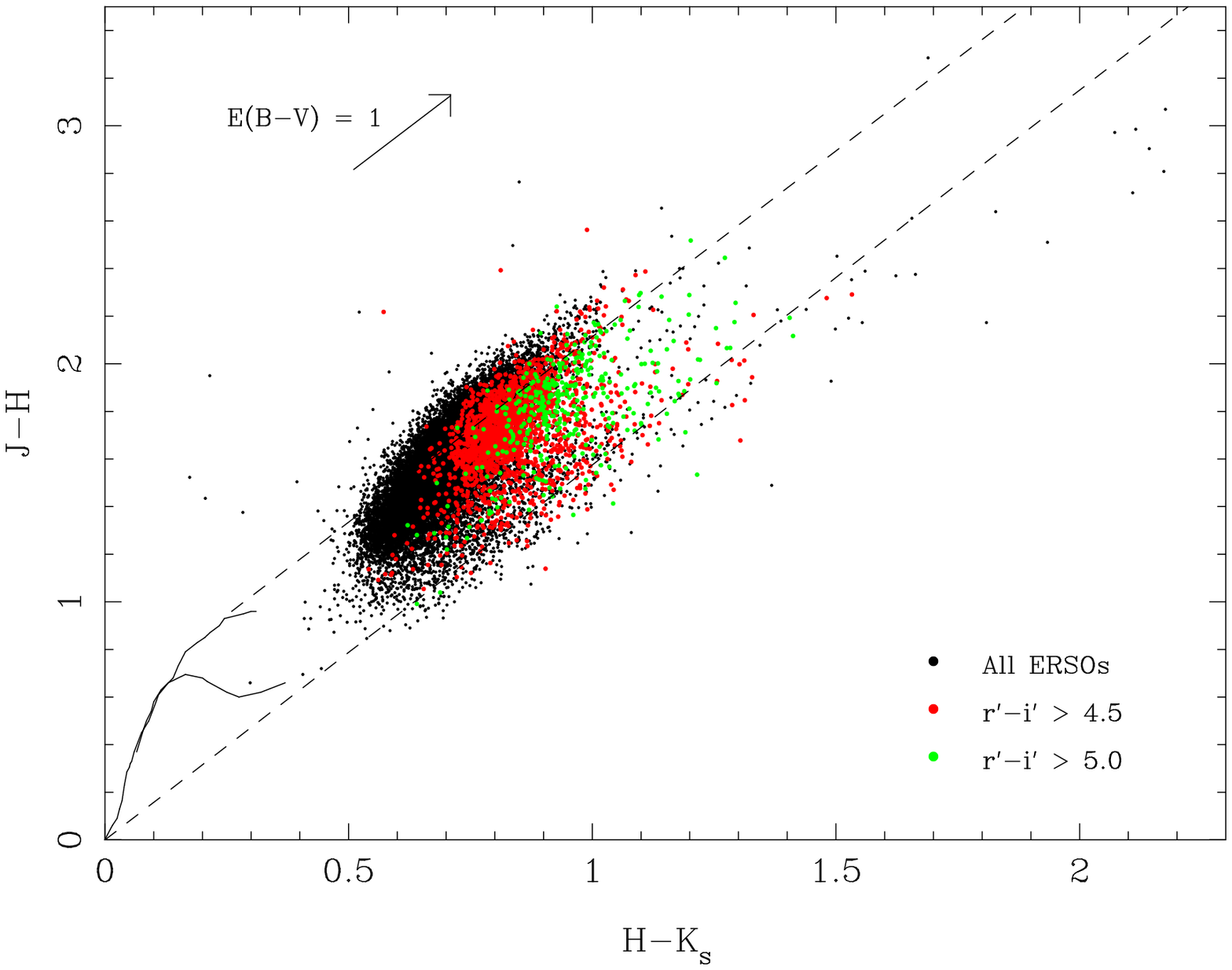}
\caption{2MASS colour-colour diagram for all ERSOs with valid quality detections (grades A-D) in all three near-IR bands, with ($r'-i'$) colours indicated. Tracks for the unreddened main-sequence to M6V (lower black line) and giant branch to M7III (upper black line), from \citet{bess88}, are shown. A reddening strips are shown as dashed lines extending away from the unreddened positions using an R$_V$~=~3.1 extinction law. An E(B-V)~=~1 reddening vector is also shown.}
\label{eros_2mass}
\end{center}
\end{figure*}

Near-IR photometry of the ERSO sample will be less affected by extinction than the optical photometry and may also reveal the effects of different chemistries in the photosphere. From the 2MASS database \citep{skru06}, $J$~(1.25~$\mu$m), $H$~(1.65~$\mu$m) and $K_s$~(2.17~$\mu$m) magnitudes were obtained for the ERSOs. Out of 25473 IPHAS ERSOs, 25411 (99.8\%) have 2MASS counterparts within 0.5\arcs (to be expected since the two databases are calibrated on the same astrometric system). Of those, 24767 (97.2\% of the ERSO sample) had valid detections in all three 2MASS bands (at A-grade level). Figure~\ref{eros_2mass} shows the 2MASS colour-colour diagram for these objects.

Also shown in Figure~\ref{eros_2mass} are the tracks for unreddened main-sequence and giant stars from \citet{bess88}, along with a reddening strip extending up and to the right defined using an $R_V = 3.1$ extinction law.  The vast majority of ERSOs fall in the area characteristic of reddened late-type giants: the spread indicates a range of reddenings, $1 < E(B-V) < 5$. The data in Figure~\ref{eros_2mass} are colour-coded to show that sources with higher $(r' - i')$ colour generally have redder
2MASS colours, in line with expectation.  Nevertheless, many of the reddest 2MASS sources are not among the reddest IPHAS sources.  This will be due to the effects of either warm circumstellar dust, or particularly cool (not-so-severely reddened) photospheres.

\begin{figure}
\begin{center}
\includegraphics[width=175pt, angle=270]{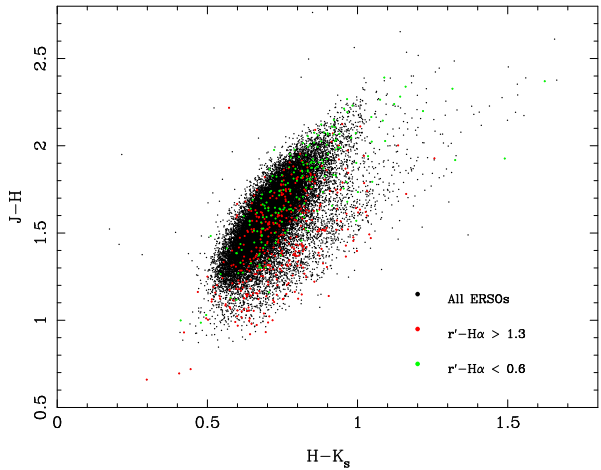}
\caption{2MASS colour-colour diagrams for all ERSOs as per Figure~\ref{eros_2mass} with the positions of objects with particularly high or low $r'$-H$\alpha$ colours indicated.}
\label{eros_double_2mass}
\end{center}
\end{figure}

In the $(J-H, H-K_s)$ plane, oxygen-rich long period variables (LPVs) are expected to lie below the normal giant reddening line, at redder $(H - K_s)$ colours, on account of their cooler photospheres and TiO, VO, CO and H$2$O blanketing. Carbon stars have been found to lie more nearly along the giant reddening line, being generally redder in both $(J-H)$ and $(H-K_s)$, due to CO, CN and C$_2$ blanketing \citep[see e.g.,][, Figure A3]{bess88}. Figure~\ref{eros_double_2mass} shows this plane with the ERSOs colour-coded according to their $(r' -H\alpha)$ index. Objects with low ($r' - $H$\alpha$) colour appear more clustered towards higher $(J-H)$ and $(H-K_s)$ colours, unlike the distribution of objects with high $(r' - $H$\alpha$) colour, which appear more evenly distributed. If carbon stars do have lower ($r' - $H$\alpha$) colours as was indicated in Section~\ref{s-simbad}, they would be expected to be found at redder colours in the 2MASS colour-colour diagram, as indeed they seem to be.

\subsection{MSX and the fraction of ERSOs with infrared excesses}
\label{s-msx}

In Section~\ref{s-galactic} it was shown that the prevalence of ERSOs is strongly correlated with galactic extinction and that therefore their highly reddened colours are likely to be dominated by it. However, given the evidence in Section~\ref{s-colours} that these sources are feasibly late-type giant stars, it is probable that many have circumstellar material which further reddens them. We turn to the MSX survey to assess the fraction of objects with infrared excesses signaling the presence of circumstellar matter.

The MSX survey covers the entire galactic plane at four mid-infrared wavelength, but due to the much higher sensitivity of the 8.28~$\mu$m band compared to the other three bands, the majority of the 300,000 objects in the MSX point source catalogue are only detected in the 8.28~$\mu$m band. There are 9371 ERSOs with valid detections in the 2MASS $K_s$ band and "fair" or better quality detections in the MSX 8.28~$\mu$m band. Hence we use the $(K_s - \mathrm{[8]})$ colour to assess the IR excess beyond that expected for purely photospheric emission \citep[which, for mid-M giants typical of the ERSO sample is $(K_s - \mathrm{[8]}) \sim 0.2$][]{pric04}.

\begin{figure}
\begin{center}
\includegraphics[width=225pt, angle=270]{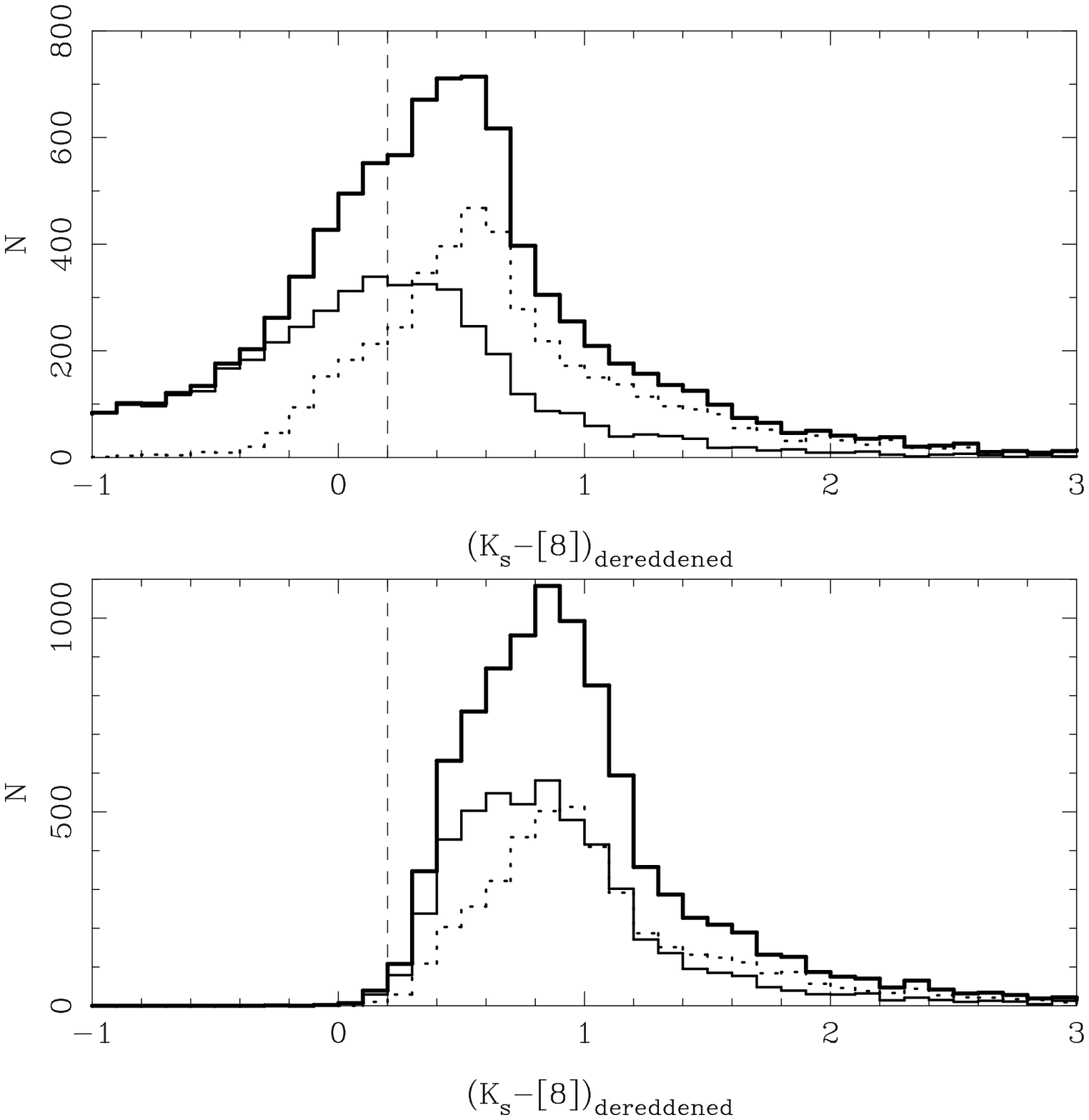}
\caption{Dereddened $(K_s - \mathrm{[8]})$ colours for all ERSOs with detections in the 2MASS $K_s$ band and the MSX 8.28~$\mu$m band (see Figure~\ref{msx_histogram}) (full line). Also shown are the $(K_s - \mathrm{[8]})$ colours for sources at low ($|b| < 1.5$, full line) and high ($|b| > 1.5$, dotted line) galactic latitudes. The data in the upper figure were de-reddened using galactic extinction data of \citet{schl98} while the data in the lower figure were de-reddened using the 1.5~kpc extinction data from \citet{mars06}. The dashed vertical line indicates the expected photospheric colour of a mid-M giant \citep{pric04}.}
\label{K8_excesses}
\end{center}
\end{figure}

As shown in Section~\ref{s-colours}, these sources are typically experiencing 6-18 magnitudes of visual extinction, or $\sim 0.5-2$ magnitudes of extinction in the $K_s$ band. Therefore, to study their $(K_s - \mathrm{[8]})$ colour we must deredden both the $K_s$ and 8~$\mu$m observations. If every object is presumed to be maximally reddened for its line of sight, they can be dereddened using the total galactic extinction data from \citet{schl98} and the 8~$\mu$m relative extinction from \citet{inde05}. Figure~\ref{K8_excesses} shows the dereddened $(K_s - \mathrm{[8]})$ colour for the 9371 ERSOs with valid detections in both bands, which peaks around $(K_s - \mathrm{[8]}) \sim 0.5$.

However, the total galactic extinction data presented by \citet{schl98} will likely over-estimate the amount of extinction for sources at low galactic latitudes which would not be expected to lie exterior to the entire Galactic dust column. In Figure~\ref{K8_excesses} we also show the colours for two parts of the ERSO sample, those at low galactic latitudes ($|b| < 1.5$) and those at high galactic latitudes ($|b| > 1.5$). The marked difference in the distribution of colours between these two samples makes sense as a consequence of the over-estimated extinction, particularly for sources at low latitudes.

A method to overcome this anomaly is to use interstellar extinction data evaluated at a finite distance instead of the asymptotic value. We dereddened the 2MASS $K_s$-band data using extinction data from \citet{mars06} evaluated at various distances with intervals of 0.25~kpc in the range of $0.5 - 5.0$~kpc. To determine which set of extinction data would best approximate that experienced by the ERSO sample we looked for that which caused the distribution of colours for the low and high galactic latitude ERSO samples to be most coincident. We determined that the most satisfactory overall match could be obtained with extinction data evaluated at 1.5~kpc, a similar distance to that at which the galactic extinction map most closely resembles the ERSO distribution (see Figure~\ref{inner_galactic}). The distribution of $(K_s - \mathrm{[8]})$ dereddened colours for this extinction limit are shown in the bottom panel of Figure~\ref{K8_excesses}. The peak has now shifted to $(K_s - \mathrm{[8]}) \sim 0.8$ for both the low and high latitude distributions.

Applying this dereddening, of the 9371 ERSOs with valid 2MASS and MSX detections, 2156 (23\%) have $(K_s - \mathrm{[8]})$ colours greater than 1.2 which form a tail extending to redder colours and most likely includes many objects with large amounts of circumstellar material. Also of note when comparing the colour distribution between the two de-reddening methods is that the FWHM has decreased from $\Delta (K_s - \mathrm{[8]}) \sim 1.2$ to $\Delta (K_s - \mathrm{[8]}) \sim 0.7$, which we interpret as evidence that choosing to limit the extinction to the value out to 1.5~kpc is an improvement. Since many of these sources are likely to be thermally-pulsating AGB stars they will experience variability in the $K_s$-band. While this will affect the spread in colours it should not affect its peak position as we should be sampling sources at all stages in the pulsation cycle. Therefore the typical colour excess, E($K_s - \mathrm{[8]})$, beyond that due to photospheric emission, is therefore $\sim$0.5 for these sources.

\begin{figure}
\begin{center}
\includegraphics[width=180pt, angle=270]{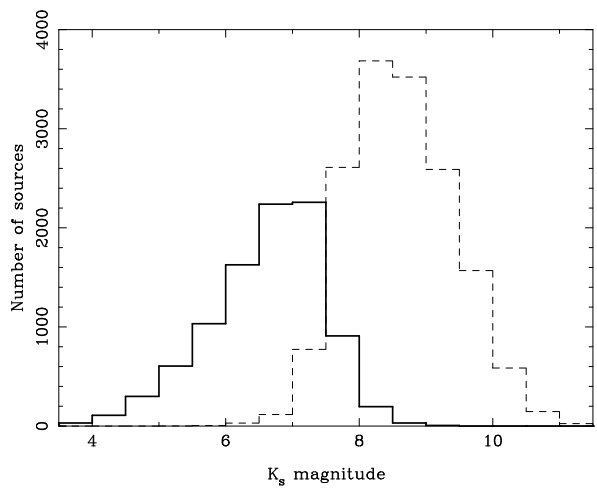}
\caption{Numbers of ERSOs with (thick black line) and without (dashed line) MSX 8.28~$\mu$m fluxes, binned into half-magnitude bins from the 2MASS $K_s$ band.}
\label{msx_histogram}
\end{center}
\end{figure}

The colour distributions in the lower panel of Figure~\ref{K8_excesses} are strong evidence for circumstellar material around a large fraction of our ERSOs with MSX detections. But we have to consider whether this is due only to including sources with detections at 8.28~$\mu$m in the first place. Figure~\ref{msx_histogram} shows the number of ERSOs with and without MSX 8.28~$\mu$m detections as a function of their 2MASS $K_s$ magnitude. The detection limit of the MSX survey can clearly be seen as a function of the near-infrared $K_s$ magnitude for the ERSOs. 90\% of sources brighter than $K_s = 7.5$~mag have an MSX 8.28~$\mu$m detection, compared to 7\% of sources fainter than $K_s = 7.5$. Therefore, it appears that the MSX-detected ERSOs are a $K_s$-bright subset, selected not on the size of their infrared excess, but on their predominantly photospheric brightness. This implies that the colour excesses of the ERSOs determined from Figure~\ref{K8_excesses} are not a product of their infrared brightness, but are likely to be true for the entire sample.

Some of the $K_s$-bright ERSOs ($K_s < 7.5$) without an 8.28~$\mu$m detection may have circumstellar material that is too cold to emit at 8~$\mu$m but is detectable at longer wavelengths. For these 1041 sources, we also searched for detections at longer wavelengths (e.g. ISOGAL and IRAS) and from deeper surveys (e.g. GLIMPSE) and found 327 ERSOs, mainly in the IRAS 60 and 100~$\mu$m bands. The lower sensitivity of long wavelength photometry makes the importance of this difficult to quantify, but there are clearly many objects with cold circumstellar dust that have escaped detection by MSX. There may be a small fraction of sources within our ERSO sample that have little or no circumstellar material, in which case their extremely red colours must be entirely due to interstellar reddening. Based on the distribution of $(K_s - \mathrm{[8]})$ colours in Figure~\ref{K8_excesses} and the fraction of $K_s$-bright ERSOs with no mid- or far-IR detections, we estimate that this fraction is at most $\sim$15\%.

In conclusion, while the extreme red colours of ERSOs are due in large part to interstellar extinction, it also appears to be the case that many of them also exhibit infrared excesses indicative of substantial circumstellar material.

\subsection{IRAS}

Here we examine the mid- to far- infrared characteristics of the ERSO sample by cross-matching to the IRAS Point Source Catalogue \citep{beic88}. Due to the lower spatial resolution of the IRAS survey \citep{neug84}, it was appropriate to use a cross-match radius of 5\arcs. 2884 (11.3\%) ERSOs were found to have detections in one or more  of the four IRAS bands at 12, 25, 60, or 100~$\mu$m. The number of objects with "high" or "moderate" quality detections increases towards shorter wavelengths, and the number of ERSO sources which match this requirement in the 12, 25, and 60~$\mu$m bands is only 206. This low fraction of matches compared to that of 2MASS is a consequence of both the much brighter IRAS limiting magnitude of m$_{12} \sim 6$ and confusion in the galactic plane. However, the cross-matches are of interest as it allows us to draw on the existing understanding of IRAS colour-colour diagrams: in particular we make use of the discussion by \citet{vand88}.

\begin{figure}
\begin{center}
\includegraphics[width=180pt, angle=270]{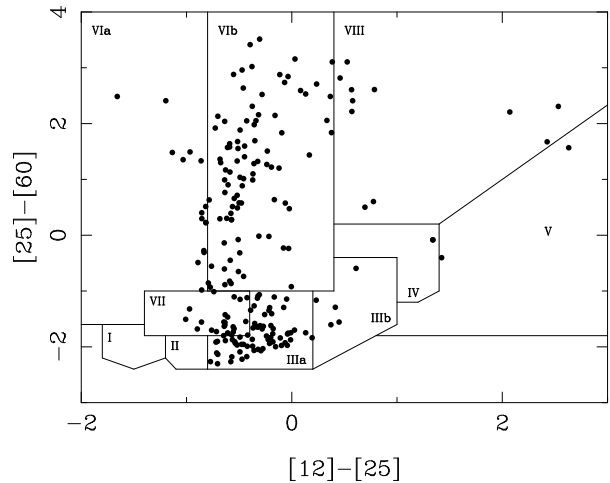}
\caption{IRAS two-colour diagram for all ERSOs with "high" or "moderate" quality detections in the 12, 25, and 60~$\mu$m IRAS bands. The colours are not colour-corrected and are defined in the text. The divisions represent the regions identified by \citet{vand88}, each of which is characterised by a different type of circumstellar material or environment.}
\label{eros_iras}
\end{center}
\end{figure}

The photospheric contribution to emission in the IRAS bands is small, with the properties of circumstellar dust dominating the observed colours. \citet{vand88} used the infrared colours, [12]-[25] and [25]-[60], defined as

\begin{equation}
[12]-[25] = 2.5 \, \mathrm{log} \left( \frac{F_{25 \mu\mathrm{m}}}{F_{12 \mu\mathrm{m}}} \right) \nonumber
\end{equation}
\begin{equation}
[25]-[60] = 2.5 \, \mathrm{log} \left( \frac{F_{60 \mu\mathrm{m}}}{F_{25 \mu\mathrm{m}}} \right) \nonumber
\end{equation}

\noindent to show that a large fraction of evolved IRAS sources fall along a band of increasing log(F$_{12\mu\mathrm{m}}$/F$_{25\mu\mathrm{m}}$), which represents an evolutionary sequence for M and C stars. Figure~\ref{eros_iras} shows our reproduction of the IRAS two-colour diagram for all ERSOs with "high" or "moderate" quality in the 12, 25, and 60~$\mu$m bands. We also show the regions identified by \citet{vand88} that allowed separation of different types of mass-losing star. In the following discussion we compare the fraction of ERSOs in each of these regions to the fraction of sources in those regions in both the total all-sky IRAS catalogue of point sources and just those in the area of the northern galactic plane covered by IPHAS.

Of note is the complete lack of ERSOs in regions I and II, which \citet{vand88} identified as containing objects with little or no circumstellar material. While the complete IRAS Catalog contains $\sim 11.4$\% of sources in these regions, the fraction is much lower in the galactic plane, with only $\sim 2.7$\% of sources. One possibility for their absence is that the ERSO sample only includes sources with significant circumstellar material. However, another possibility is that our requirement of $(r' - i') > 3.5$ has feasibly excluded these objects because they are typically less evolved than the mass-losing AGB stars and are likely to have intrinsically bluer colours because of their earlier spectral type. Sources with purely photospheric emission will be particularly faint in the 25 and 60~$\mu$m IRAS bands and the majority may not have been detected in these bands.

The low fractions of objects in regions V and VIII can also be attributed to our selection criteria, which will exclude planetary nebulae and early-type stars due to their bluer optical colours and extended morphology which will be classified as non-stellar sources in the IPHAS catalogue \citep{gonz08}.

The fractions of sources in the remaining six regions are shown in Table~\ref{iras_regions2} compared to the fractions of IRAS source is those regions from the all-sky catalogue and those in the area of the northern galactic plane. The fractions in these regions are very similar for the three samples. The largest difference is between the regions identified by \citet{vand88} as containing significant fractions of C-rich objects, regions VIa and VII, and the regions more associated with O-rich sources, regions IIIa and VIb. The ERSO sample appears to show a higher fraction of O-rich objects than the all-sky IRAS sample, as might be expected for a sample drawn from the higher metallicity galactic disk compared to a sample containing objects from the lower-metallicity halo. However, this trend is not as clear in the sample of galactic plane IRAS sources, throwing doubt on this explanation.

The high fraction of 60~$\mu$m excess sources in region VIb of the IPHAS sample is echoed in the IRAS NGP sample. The 60~$\mu$m excess is believed in be indicative of a detached circumstellar shell which might either be due to an interruption to the mass-loss process \citep{zijl92} or due to swept-up ISM material \citep[e.g.][]{ware07}. However, the higher fraction of 60~$\mu$m excess source in the galactic plane indicates that source confusion and contributions from background emission may be influencing the colours of these samples. Inspection of the IRAS images confirms that this is likely to be responsible for the colours of $\sim$10-20\% of the sources in our sample, which might bring the fraction more in line with that of the IRAS all-sky sample.

The low fractions of objects in regions IIIb and IV (both associated with sources with generally thicker circumstellar shells) for the ERSO sample, while not statistically significant given the small number of sources in those regions, could be attributed to the difficulty in detecting highly obscured sources in any optical survey.

Of the 4653 IRAS-identified evolved stars in the area of the northern galactic plane covered by IPHAS, only 206 have been associated with ERSOs. If the remaining IRAS sources have optical association in the IPHAS catalogue, they are likely to belong to less optically-reddened sources, despite their strong infrared emission. The similar fractions of objects in the six IRAS regions in Table~\ref{iras_regions2} indicates that the ERSOs do not represent an extreme subset of the overall IRAS sample in terms of circumstellar emission, and therefore may not necessarily represent the IPHAS sources with the strongest circumstellar excesses.

Considering the evidence in Section~\ref{s-galactic} that the ERSOs are a sample dominated by the effects of interstellar reddening, the subset of IRAS-detected ERSOs represent the ERSOs with the brightest circumstellar material (as would be necessary for a detection with IRAS). Whereas as a subset of the overall IRAS catalogue, the IRAS-detected ERSOs are those IRAS sources experiencing the strongest interstellar reddening.

\begin{table}
\caption{Star counts for ERSOs in the different regions of the IRAS two-colour diagram, as specified by \citet{vand88}. These are compared with the fractions of such objects in both the complete all-sky IRAS catalogue of point sources and all IRAS sources in the area of the northern galactic plane (NGP) covered by IPHAS. Column~2 lists typical properties of the objects found in the region. We show only counts for regions IIIa, IIIb, IV, VIa, VIb and VII, since the other regions are not found to contribute to the ERSO sample due to our imposed selection criteria.}
\begin{tabular}{@{}lrrrrrr}
\hline
Region & \multicolumn{2}{c}{IRAS All-sky} & \multicolumn{2}{c}{IRAS NGP} & \multicolumn{2}{c}{ERSOs}\\
\cline{2-3} \cline{4-5} \cline{6-7} \\
 & No. & \% & No. & \% & No. & \% \\
\hline
IIIa 			& 3215	& 25.8	& 516	& 17.2 	& 66 		& 33.8\\
IIIb 			& 675	& 5.4		& 122	& 4.1		& 5		& 2.6\\
IV 			& 258	& 2.1		& 44		& 1.5		& 2		& 1.0\\
VIa 			& 1329 	& 10.7 	& 350	& 11.7	& 15		& 7.7\\
VIb 			& 4228	& 33.8	& 1410	& 47.1	& 84		& 43.1\\
VII 			& 2769	& 22.2	& 551	& 18.4	& 23		& 11.8\\
\hline
Totals: 	& 12474	& 100	& 2993	& 100	& 195 & 100\\
\hline
\end{tabular}
\label{iras_regions2}
\end{table}

\section{Spectroscopic follow-up observations}
\label{s-spectra}

\begin{figure}
\begin{center}
\includegraphics[height=240pt, angle=270]{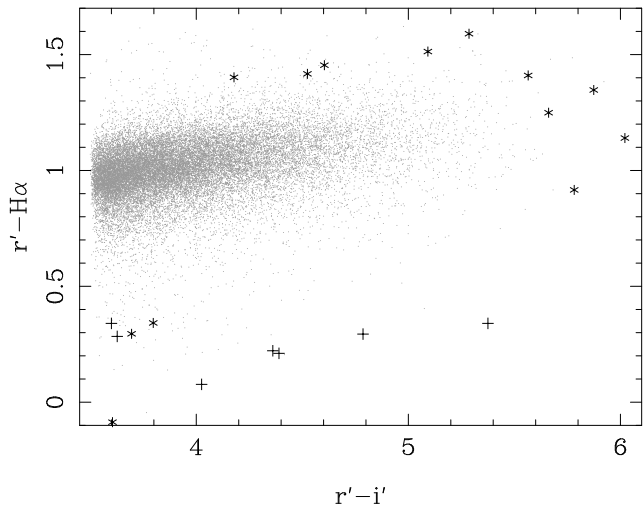}
\caption{The extremely red region of the IPHAS colour-colour plane showing all ERSOs as grey dots. The positions of all sources for which spectra were taken are shown: star symbols are O-rich stars and plus symbols indicate S-type stars. See the text for discussion of derived spectral types.}
\label{spectra_colours}
\end{center}
\end{figure}

There are many objects in the IPHAS colour-colour plot in Figure~\ref{ERSO_colcol} with colours so extreme that they warrant further investigation. Of particular note are the large numbers of objects with $(r' - $H$\alpha)$ values significantly below the main strip at $(r' - $H$\alpha) \sim 1$ whose nature is not immediately obvious, but which may include some sources with atypical surface chemistries (see Section~\ref{s-simbad} for sources already discovered in this area). This section comprises a series of spectroscopic follow-up observations to determine if the extreme objects in the ERSO sample exhibit any particularly notable properties which may allow interesting types of object to be readily extracted from the IPHAS catalogue. The positions in the IPHAS colour-colour plane for the objects discussed in this section are shown in Figure~\ref{spectra_colours}.

These observations will be discussed more fully in a future paper, but we summarise the observational procedure here. The observations were carried out using LIRIS \citep[Long-slit Intermediate Resolution Infrared Spectrograph,][]{acos03}, a near-IR spectrograph on the William Herschel Telescope (WHT) on five nights during the summers of 2006 and 2007 in generally good conditions. Each target was observed using both the lrzj8 ($0.89 - 1.51 \mu$m) and lrhk ($1.40 - 2.39 \mu$m) grisms, giving continuous coverage from 0.89-2.39~$\mu$m through the z, J, H and K bands. The resolving power was 700 in both grisms. Each target was observed twice at different positions along the slit and the sky observations subtracted to remove the background. Wavelength calibration was performed using argon and xenon lamps available at the telescope. Telluric correction was performed using observations of infrared standard stars made throughout each night. In some regions, where atmospheric transmission was near zero no useful information could be reconstructed. Finally a continuum was estimated for each object and the spectrum was divided by this.

The near-IR region was chosen partly because of the high infrared brightness of these sources and also because of the wide range of molecular and atomic lines visible in the infrared. Molecular bands due to TiO, VO, ZrO, and C$_2$ appear through the $z$ and $J$ bands and allow clear chemical classes to be identified and spectral types assigned. The CO bands in the $K$ region have been shown to be good indicators of luminosity \citep{lanc07} and the wide near-IR H$_2$O bands are excellent indicators of variability in evolved stars \citep{lanc00}. Unless otherwise stated, errors on the derived temperature classes are $\pm 1$ type.

\subsection{Objects with low $r' - $H$\alpha$ colours}
\label{s-lowrha}

\begin{figure}
\begin{center}
\includegraphics[height=240pt, angle=270]{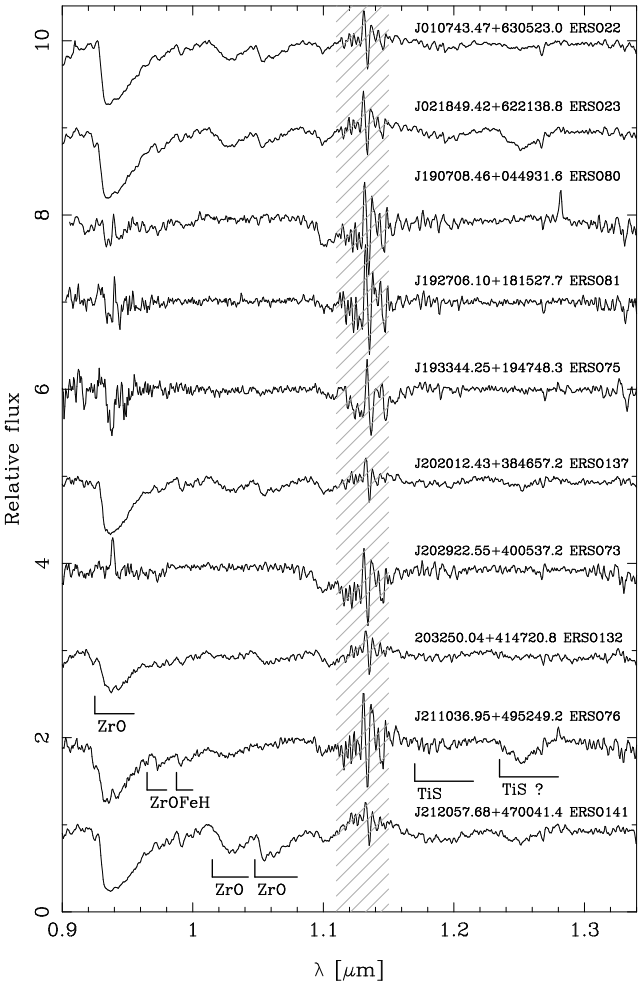}
\caption{zJ grism spectra of objects with low $r' - $H$\alpha$ colours, as listed in Table~\ref{objects_lowrha}. Each spectrum has been corrected for telluric absorption and been divided by an adopted continuum. The spectra have been separated by integer values of normalised flux to make each clear and visible. The shaded region indicates a region of low atmospheric transmission where our telluric correction was unable to recover a useful spectrum. Prominent molecular features have been marked.}
\label{spectra_lowrha}
\end{center}
\end{figure}

The spread of $(r' - $H$\alpha)$ colours below the main band in Figure~\ref{ERSO_colcol} is quite large, with colours extending down to $(r' - $H$\alpha) \sim 0$, well below the early~A reddening line identified by \citet{drew05} as the lowest expected position of non-degenerate stars with H$\alpha$ absorption. In Section~\ref{s-simbad} a small number of previously identified carbon and S-type stars were shown to have colours putting them in this area and we might therefore expect to find more of these types of object in this sample. We have identified 61 objects with $(r' - $H$\alpha) < 0.4$ in the ERSO region of the IPHAS colour-colour plane, significantly below the expected position of the early-A reddening line at $(r' - $H$\alpha) \sim 0.55-0.60$ for $(r' - i') = 3.5 - 6.0$ \citep[the reddened positions of early-A dwarfs, below which no normal stars should lie,][]{drew05}. We have obtained near-IR spectra of 10~targets from this group, including one object with $(r' - $H$\alpha) = -0.086$, the details of which are listed in Table~\ref{objects_lowrha}, including approximate spectral types derived from the observations. zJ grism spectra of all objects are shown in Figure~\ref{spectra_lowrha}.

\begin{table*}
\caption{ERSOs with ($r' - $H$\alpha) < 0.4$ observed in the near-IR with LIRIS on the WHT. IPHAS 
photometry is the mean of two or more observations depending on the need for re-observations. The spectra are shown in Figure~\ref{spectra_lowrha}.}
\begin{tabular}{@{}llccccccccl}
\hline
ERSO & Name & RA (J2000) & Dec (J2000) & \multicolumn{3}{c}{IPHAS photometry} & \multicolumn{3}{c}{2MASS photometry} & Spectral \\
No. & & & & $r'$ & $r' - i'$ & $r' - $H$\alpha$ & $J$ & $H$ & $K_s$ & type\\
\hline
22 & IPHAS J010743.47+630523.0 & 01:07:43.47 & +63:05:23.0 & 17.071 & 3.600 & 0.340 & 8.001 & 6.555 & 5.899 & SX/6 \\
23 & IPHAS J021849.42+622138.8 & 02:18:49.42 & +62:21:38.8 & 17.996 & 3.627 & 0.384 & 9.143 & 7.813 & 7.203 & SX/6 \\
80 & IPHAS~J190708.46+044931.6 & 19:07:08.46 & +04:49:31.6 & 18.617 & 4.786 & 0.294 & 7.783 & 6.213 & 5.217 & SC9/8e \\
81 & IPHAS J192706.10+181527.7 & 19:27:06.10 & +18:15:27.7 & 21.212 & 3.695 & 0.295 & 12.495 & 10.698 & 9.950 & K-M {\sc iii} \\
75 & IPHAS J193344.25+194748.3 & 19:33:44.25 & +19:47:48.3 & 20.348 & 3.797 & 0.342 & 11.338 & 9.590 & 8.814 & K-M {\sc iii} \\
137 & IPHAS J202012.43+384657.2 & 20:20:12.43 & +38:46:57.2 & 19.816 & 4.390 & 0.211 & 9.236 & 7.209 & 6.284 & SX/6 \\
73 & IPHAS J202922.55+400537.2 & 20:29:22.55 & +40:05:37.2 & 19.790 & 3.605 & -0.086 & 10.074 & 8.028 & 7.049 & K-M {\sc iii} \\
132 & IPHAS J203250.04+414720.8 & 20:32:50.04 & +41:47:20.8 & 21.806 & 5.375 & 0.340 & 9.676 & 7.511 & 6.465 & SX/7 \\
76 & IPHAS J211036.95+495249.2 & 21:10:36.95 & +49:52:49.2 & 18.912 & 4.025 & 0.077 & 10.216 & 8.461 & 7.507 & SX/7e \\
141 & IPHAS J212057.68+470041.4 & 21:20:57.68 & +47:00:41.4 & 20.293 & 4.361 & 0.222 & 8.794 & 7.444 & 6.783 & SX/6 \\
\hline
\end{tabular}
\label{objects_lowrha}
\end{table*}

Of immediate note is the high proportion of objects in this group which display the characteristic ZrO bands of the S-type stars. The strongest ZrO band visible in our spectra is at 0.93-0.96~$\mu$m \citep[the $b'^3\Pi-a^3\Delta$ system 0-0 bandhead,][]{phil79}, its distinctive shape and depth allowing it to be distinguished from TiO bands of the $\epsilon$-system \citep[e.g.][]{dank94, schi99}. However the most unmistakable ZrO bands are the pair at 1.03 and 1.06~$\mu$m \citep[the $a^3\Delta$~0-1 and $B^1\Pi-A^1\Delta$~0-0 band heads respectively,][]{hink89}. We find that six of the ten stars observed show one or more of these features and we thus identity ERSOs 22, 23, 76, 132, 137 and 141 as S-type stars. Many of these also show the 0.974~$\mu$m ZrO feature \citep[the head of the B-A $\Delta\nu = -1$ band,][]{davi81} and the 0.99~$\mu$m FeH band \citep[the $^4\Delta-^4\Delta$ 0-0 head,][]{lamb80}, both visible in the long-wavelength wing of the strong 0.93~$\mu$m ZrO feature (Figure~\ref{spectra_lowrha}).

The classification of S-type stars is based on two parameters, their temperature type and abundance index \citep{keen80}. The temperature type mirrors that for oxygen-rich stars, with values from 0-10 spanning surface temperatures of 4000-2000~K, while the abundance index varies from 1-10 with increasing strength of ZrO over TiO. While this is also a good indicator of the C/O ratio, \citet{garc07} showed that the strength of the ZrO bands is often more dependent on the zirconium abundance and \citet{zijl04} found that changes to the photospheric temperature can cause significant variation in the molecular bands observed which may affect the estimated C/O ratio. Complications involving variable Zr or Ti abundances, the C/O ratio and the presence of sulphides rather than oxides further complicate matters. \citet{keen80} based their classification system on molecular bands in the optical, both for the abundance index and the temperature type and little work since then has been done in the near-infrared. \citet{joyc98b} presented $J$-band spectra of some S-type stars and we have used their work in combination with the guidelines given by \citet{keen80} to classify these objects.

Of use is the currently unidentified feature at 1.25~$\mu$m which \citet{joyc98b} observed in the spectra  of many S-type stars, all of which had a high abundance index of 5-7. This coincides with the positions of the ZrS b$'$-a $\Delta\nu = 0$ sequence and the TiS A-X $\Delta\nu = 0$ sequence, both of which may contribute to this feature. ERSOs 23, 76 and 141 show prominent features in this area, with ERSOs 23 and 141 also showing evidence for the TiS A-X $\Delta\nu = -1$ sequence at 1.17-1.22~$\mu$m, an association which may point towards the origin of the feature. The three other S-type stars show minor evidence for small features in this area when compared to spectra of non S-type stars, and the absence of any features may be related to the sulphur abundance in these objects instead of just the abundance or temperature index.

None of the S-type stars show evidence for clear TiO or VO bands, though many are coincident with the ZrO bands, making them hard to clearly identify. The 1.104~$\mu$m TiO band is in a relatively clear region of the spectrum, despite being on the edge of a large telluric feature. It sits alongside the CN red bands \citep[1.088-1.097~$\mu$m,][]{hink89} and the 1.108~$\mu$m TiS band and the three contributions are easily separable. There is no evidence for TiO bands in this region and this puts strong constraints on the abundance class of these objects. We assign spectral types of either SX/6 (ZrO strong, no TiO) or SX/7 \citep[ZrO weaker, no TiO][]{keen80} to these objects, as types listed in Table~\ref{objects_lowrha}. All these objects have C/O ratios very close to unity and the metal oxide bands vary very weakly with temperature \citep{keen80} so that ratios of atomic lines are the preferred method for assigning temperature class. Since there are not any identified or calibrated atomic lines ratios in the infrared we are unable to assign temperature classes for these objects and leave them with the symbol X to mark the unknown temperature class.

One of the objects in our sample, ERSO-80, was classified by \citet{step90} as an S-type star, noting that the object had no TiO bands, weak LaO bands and was quite red. Our spectra do not show any TiO or ZrO features, but do show moderate CN bands at 1.088~$\mu$m, as well as a very clean $H$-band CO spectrum with no OH bands, both indicative of a high C/O ratio. The presence of weak LaO bands combined with CN lines supports this. There is no evidence for any of the bands of C$_2$ which often appear for C/O$ > 1$, so our spectral classification is limited to an abundance class of 8, which \citet{keen80} define as stars having no ZrO or C$_2$, and C/O~$\sim 1$. The deep CO bands and the presence of LaO, which is only thought to be visible for T$_{eff} < 2800$~K, indicates a relatively cool photosphere and a temperature class around $9 \pm 2$. Its IRAS colours put it in either region IIIa or VIb in Figure~\ref{eros_iras} (depending on the exact value of the 60~$\mu$m flux for which there is only an upper limit available), indicating circumstellar material that is probably O-rich, as would be expected for an object that has been O-rich for most of its life and is only now on the transition to becoming C-rich. The Paschen~$\beta$ line is present in emission, indicating it may be a Mira variable.

The remaining three stars in Figure~\ref{spectra_lowrha} are relatively devoid of useful features for diagnosing their spectral types. There is no evidence for either the ZrO or TiO features around 0.93~$\mu$m and the effects of telluric correction have left the area very noisy. All three sources show CO and OH bands in the $H$-band which suggests oxygen-rich surface chemistries, while the lack of TiO or VO features imply that they are too warm for the molecules to survive \citep[TiO bands in the infrared develop only for types later than M2,][]{lanc07}. Since CO bands only develop at temperatures significantly below $\sim 5000$~K \citep{lanc07}, the spectral types of these objects are around K0-M2.

Out of ten objects with $r' - $H$\alpha < 0.4$, seven show evidence of being S-type stars. A possible explanation for the unusually low ($r' - $H$\alpha$) colours of the three O-rich giants is that because they are relatively faint their measurements may have large photometric errors (ERSOs 75 and 81 are fainter than the 10$\sigma$ detection threshold in the $r'$ band). Despite this, it is clear that the area below the main locus in the IPHAS colour-colour plane contains many objects with atypical surface chemistry and C/O ratios close to unity.

\subsection{Objects with high $r'$-H$\alpha$ colours}

\begin{figure}
\begin{center}
\includegraphics[height=240pt, angle=270]{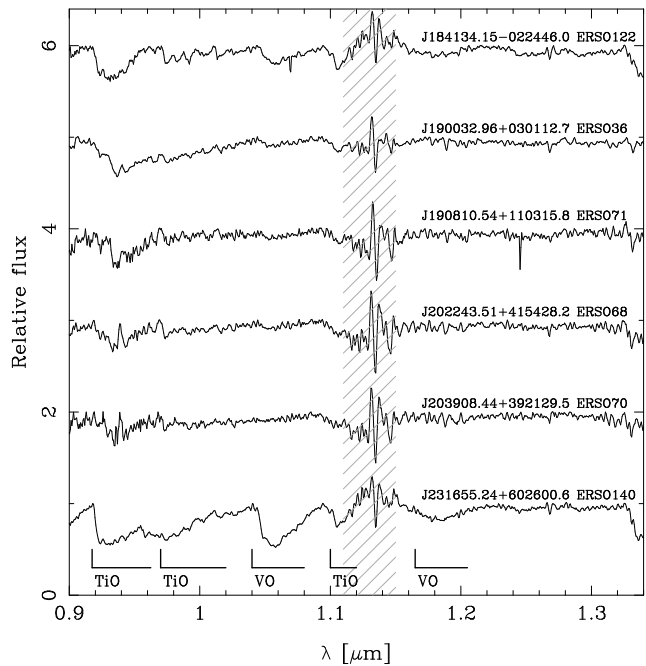}
\caption{zJ grism spectra of objects with high $r' - $H$\alpha$ colours as listed in Table~\ref{objects_highrha}. Each spectrum has been corrected for telluric absorption and has been divided by an adopted continuum. The spectra have been separated by integer values of normalised flux to make each clear and visible. The shaded area indicates a region of low atmospheric transmission where our telluric correction was unable to recover a useful spectrum. Prominent molecular features have been marked.}
\label{spectra_highrha}
\end{center}
\end{figure}

The high $(r' - $H$\alpha)$ side of the histograms in Figure~\ref{rha_mags} shows relatively little spread. There are 107 objects with $(r' - $H$\alpha) > 1.4$, the highest of which reach $(r' - $H$\alpha) \sim 1.65$. We have taken near-IR spectra of six of these objects, as listed in Table~\ref{objects_highrha} and shown in Figure~\ref{spectra_highrha}. The presence of objects with an H$\alpha$ excess in the IPHAS survey has been discussed by \citet{with08} and \citet{corr08} and the latter noted that some symbiotic stars may show IPHAS colours that would put them above the main group in the ERSO region. Mira variables can also show phase-dependent H$\alpha$ emission due to shocks in their atmosphere from stellar pulsation. If the H$\alpha$ emission was significant, it could cause a sufficient ($r' - $H$\alpha$) excess to raise the objects above the main group in the colour-colour diagram.

\begin{table*}
\caption{ERSOs with $(r' - $H$\alpha) > 1.4$ observed in the near-IR with LIRIS on the WHT as per 
Table~\ref{objects_lowrha}. Note: J231655.24+602600.6 does not have valid detections in the 2MASS $J$ and $H$ bands. Spectra for all six objects are shown in Figure~\ref{spectra_highrha}.}
\begin{tabular}{@{}llccccccccl}
\hline
No. & Name & RA (J2000) & Dec (J2000) & \multicolumn{3}{c}{IPHAS photometry} & \multicolumn{3}{c}
{2MASS photometry} & Spectral \\
 & & & & $r'$ & $r' - i'$ & $r' - $H$\alpha$ & $J$ & $H$ & $K_s$ & type \\
\hline
122 & IPHAS~J184134.15-022446.0 & 18:41:34.15 & -02:24:46.0 & 21.844 & 5.286 & 1.590 & 10.343 
& 8.805 & 8.065 & M6.5 {\sc iii} \\
36 & IPHAS~J190032.96+030112.7 & 19:00:32.96 & +03:01:12.7 & 21.899 & 5.565 & 1.410 & 9.515 & 7.391 & 6.442 & M5.5 {\sc iii} \\
71 & IPHAS~J190810.54+110315.8 & 19:08:10.54 & +11:03:15.8 & 18.108 & 4.178 & 1.402 & 8.419 & 6.883 & 6.131 & M4 {\sc iii} \\
68 & IPHAS~J202243.51+415428.2 & 20:22:43.51 & +41:54:28.2 & 21.582 & 5.092 & 1.513 & 9.801 & 7.800 & 6.863 & M4 {\sc iii} \\
70 & IPHAS~J203908.44+392129.5 & 20:39:08.44 & +39:21:29.5 & 22.445 & 4.604 & 1.454 & 11.867 & 9.843 & 8.930 & M2 {\sc iii}\\
140 & IPHAS~J231655.24+602600.6 & 23:16:55.24 & +60:26:00.6 & 16.094 & 4.524 & 1.417 & - & - & 3.227 & M8.5 {\sc iii} \\
\hline
\end{tabular}
\label{objects_highrha}
\end{table*}

Two of the objects in Figure~\ref{spectra_highrha} show clear VO features around 1.05~$\mu$m that mark them out as cool oxygen-rich giants. The VO feature only becomes visible at spectral types M6 and later \citep[e.g.][]{joyc98b} and its strength grows with decreasing temperature. ERSOs 122 and 140 are therefore easily classified based upon the strength of this feature. ERSO140 also shows very strong H$_2$O absorption in the wings of the $H$-band, a feature indicative of variability since H$_2$O may form in the cool, extended atmospheres of highly variable stars. SIMBAD lists ERSO-140 as V563~Cas, an M6 variable star. Based upon its strong VO feature we assign a spectral type of M8.5, in approximate agreement with the previous classification given the variation of spectral type expected over the pulsation cycle of an evolved star \citep[e.g.][]{lanc00}.

ERSOs 36, 71 and 68 show small depressions around 1.05~$\mu$m, which may be due to a VO feature. They all show CO and OH lines in their $H$-band spectra, so are clearly O-rich and the strengths of the $K_s$ CO bands indicates relatively cool photospheres. TiO bands that may be useful for further classification of oxygen-rich giants include the $\epsilon$-system, with the $\Delta\nu = 1$ bandhead at $\sim 0.93 \mu$m \citep{schi99}. A weak telluric feature exists at this position and for some of our observations the flux drops to nearly zero, causing any attempts to correct for this to have high levels of noise (e.g. ERSO-75 in Figure~\ref{spectra_lowrha} is a particularly bad case). In other objects where the telluric absorption effects are not as bad, or the flux levels do not drop so much, a relatively noise-less spectrum may be reconstructed (e.g. ERSO-36 in Figure~\ref{spectra_highrha}). In these cases the strength of the underlying TiO feature may be estimated, from which the stellar temperature may be estimated \citep[e.g.][]{bret90,schi99}. The combination of minor VO depressions and evidence for TiO absorption in ERSOs 36, 71 and 68 has allowed us to derive the spectral types listed in Table~\ref{objects_highrha} with an estimated error of $\pm 2$ subtypes. The final object, ERSO-70 shows no evidence for VO absorption and suffers from noisy telluric correction around 0.93~$\mu$m. The presence of OH bands in the $H$-region as well as weak TiO absorption at 1.10~$\mu$m  allows an approximate classification of M2, with an estimated error of $\pm 3$ subtypes.

All six of the objects with high $(r' - $H$\alpha)$ colours that we have spectra for are O-rich M giants of variable spectral type. We see no evidence for emission lines in the infrared (though the separation between photometric survey observations and near-IR spectroscopic observations is 1-2 years which may cause changes in some spectral features if the stars are variable). Their high $(r' - $H$\alpha)$ colours may have been due to temporary H$\alpha$ emission from circumstellar material, or particularly deep molecular features in the $r'$~band that would cause the pseudo-continuum to be lower than is usual for M-giants (see Section~\ref{s-extract} for a discussion of this effect).

\subsection{Objects with extremely high $r' - i'$ colours}

\begin{figure}
\begin{center}
\includegraphics[height=240pt, angle=270]{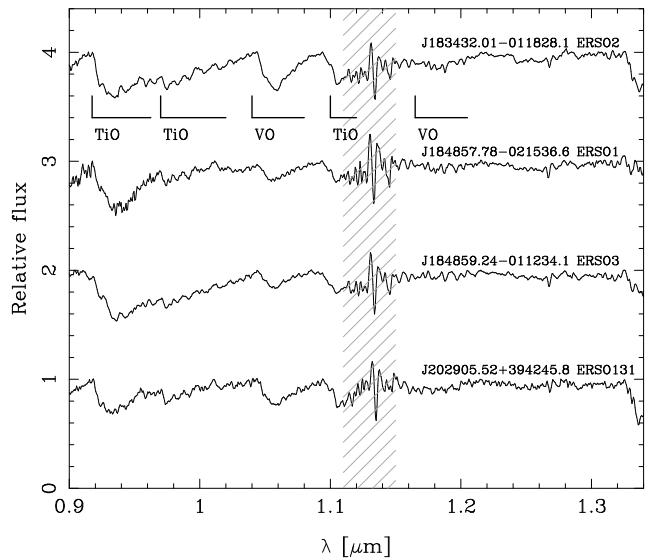}
\caption{zJ grism spectra of objects with high $(r' - i')$ colours, as listed in Table~\ref{objects_highri}. Each spectrum has been corrected for telluric absorption and has been divided by an adopted continuum. The spectra have been separated by integer values of normalised flux to make each clear and visible. The shaded area indicates a region of low atmospheric transmission where our telluric correction was unable to recover a useful spectrum. Prominent molecular features have been marked.}
\label{spectra_highri}
\end{center}
\end{figure}

While the ERSOs predominantly have $3.5 < (r' - i') < 5.0$, there are a small numbers of objects up to $(r' - i') \sim 6.0$. We have identified 19 objects with $(r' - i') > 5.5$ and have obtained near-IR spectra of 5 of them, as listed in Table~\ref{objects_highri} and shown in Figure~\ref{spectra_highri} (except ERSO-36 which is shown in Figure~\ref{spectra_highrha} since it has both high $(r' - $H$\alpha)$ and $(r' - i')$ colours).

\begin{table*}
\caption{ERSOs with $r'$~-~i$' > 5.5$ observed in the near-IR with LIRIS on the WHT as per Table~\ref
{objects_lowrha}. Spectra for all the objects are shown in Figures \ref{spectra_highrha} (ERSO36) and \ref{spectra_highri}.}
\begin{tabular}{@{}llccccccccl}
\hline
No. & Name & RA (J2000) & Dec (J2000) & \multicolumn{3}{c}{IPHAS photometry} & \multicolumn{3}{c}
{2MASS photometry} & Spectral \\
 & & & & $r'$ & $r' - i'$ & $r' - $H$\alpha$ & $J$ & $H$ & $K_s$ & type \\
\hline
2 & IPHAS J183432.01-011828.1 & 18:34:32.01 & -01:18:28.1 & 21.972 & 5.661 & 1.249 & 9.049 & 7.165 & 6.028 & M7.5 {\sc iii} \\
1 & IPHAS J184857.78-021536.6 & 18:48:57.78 & -02:15:36.6 & 21.721 & 5.782 & 0.916 & 8.350 & 6.211 & 5.134 & M6 {\sc iii} \\
3 & IPHAS J184859.24-011234.1 & 18:48:59.24 & -01:12:34.1 & 21.838 & 6.021 & 1.140 & 8.613 & 6.552 & 5.399 & M6 {\sc iii}\\
36 & IPHAS J190032.96+030112.7 & 19:00:32.96 & +03:01:12.7 & 21.899 & 5.565 & 1.410 & 9.515 & 7.391 & 6.442 & M5.5 {\sc iii}\\
131 & IPHAS J202905.52+394245.8 & 20:29:05.52 & +39:42:45.8 & 20.502 & 5.874 & 1.347 & 6.768 & 4.780 & 3.750 & M6.5 {\sc iii} \\
\hline
\end{tabular}
\label{objects_highri}
\end{table*}

Of immediate note is the presence of VO bands in the spectra of all five objects, as well as the TiO bands at 0.93 and 1.10~$\mu$m, which marks all these objects as particularly late M-giants. ERSOs 2, 3 and 131 all show particularly strong H$_2$O absorption in the wings of the $H$-band, which has been shown to be evidence for stellar variability \citep{lanc00}.

ERSOs 1 and 3 show very strong CO bands in their $H$-band with no evidence for OH features. This indicates that their C/O ratios may be close to unity, though the presence of strong O-rich features and no evidence for {\it s}-process enrichment confirms that these objects are still O-rich giants and have yet to become S-type stars (though not all M-type stars will be able to dredge up enough carbon during their AGB lifetime to make this transition - it is dependent on their initial mass). Due to the presence of VO bands, accurate spectral classification is relatively simple and the classifications in Table~\ref{objects_highri} are estimated to be accurate to $\pm 1$ spectral subtype.

The extremely red colours of these objects may partly be due to their late spectral types, but they are only slightly later than the objects in Table~\ref{objects_highrha}. The difference in $(r' - i')$ colour between an M5 and M10 giant is $\sim 1.1$ at any particular reddening \citep{drew05}, approximately equivalent to the reddening effect of E(B-V)~$\sim 1.6$, which indicates that both factors could be important in producing objects with significantly reddened colours. The infrared excesses of all five objects in Table~\ref{objects_highri} are significantly higher than the typical excesses of our overall sample of ERSOs and indicates a contribution from circumstellar material to the IPHAS colour.

We therefore propose that the reddest of the ERSOs are not necessarily later in spectral type than the rest of the objects in the region, but are suffering from considerably more reddening, both from circumstellar and interstellar material. How much of the reddening is due to circumstellar or interstellar material is hard to determine without an estimate of the distance to the source, or an estimate of the circumstellar reddening from SED fitting to the infrared photometry. The distance-dependent reddening maps of \citet{mars06} may allow the possible interstellar reddening along a sight-line to be gauged, and so allow an estimate of the amount of circumstellar reddening.

\section{Selecting chemically diverse stars from the IPHAS catalogue}
\label{s-extract}

It was shown in Section~\ref{s-lowrha} that a significant fraction of objects with ($r' - $H$\alpha$) colours below the main stellar locus are S-type stars, their spectra exhibiting the ZrO bands indicative of a near-unity surface C/O ratio. This is expected to not just be true for the extremely red objects ($r' - i' > 3.5$), but for all IPHAS sources which exhibit colours indicative of evolved late-type stars (e.g. Figure~\ref{fields_cc}). This section explains why these objects lie in a separate region of the IPHAS colour-colour plane to O-rich evolved stars and discusses a potential method for confidently extracting these objects from the IPHAS point source catalogue.

S-type stars occupy a short-lived but important evolutionary phase during the transition from O-rich to C-rich surface chemistry. The short duration of this evolutionary phase, combined with the lack of any photometric method in the optical or infrared to identify these objects contributes to the rarity. A search of the SIMBAD database reveals 938 known S-type stars, reinforcing their relative scarcity amongst evolved stars when compared to the much larger number of M-type stars known.

\subsection{The IPHAS H$\alpha$ filter as a molecular chemistry indicator}

\begin{figure}
\begin{center}
\includegraphics[width=250pt, angle=270]{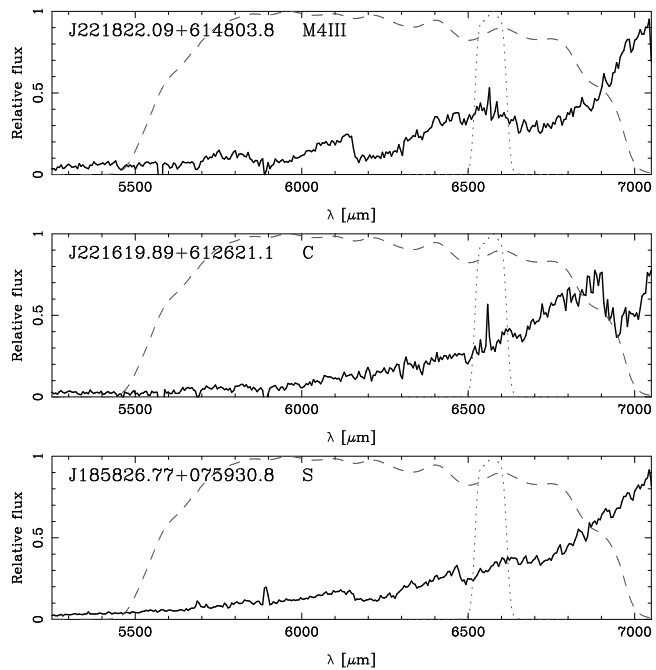}
\caption{Spectra of three evolved stars taken from the IPHAS follow-up spectroscopy programme using HectoSpec, a multi-objects spectrograph on the MMT \citep{drew05}. The $r'$ and H$\alpha$ filter profiles are shown overlaid with dashed and dotted lines respectively.}
\label{bands_compare}
\end{center}
\end{figure}

Figure~\ref{bands_compare} shows optical spectra obtained with the Multiple Mirror Telescope (MMT) HectoSpec facility for three chemically different types of cool, evolved star. These were obtained as part of a follow-up programme designed to explore the IPHAS colour-colour plane which will be described in a future paper (Steeghs et al. 2008, in preparation). \citet{drew05} provide details of the field in Cepheus from which the M and C~star HectoSpec spectra were obtained. The S-type star spectrum was taken from a later HectoSpec observation of a field in Aquila.

In the spectra of cool oxygen-rich stars the continuum within the $r'$-band is depressed by strong molecular absorption bands due to TiO and VO, however the bandpass of the narrow H$\alpha$ filter lies between these absorption features and does not experience strong depression. The result of this is that as the photospheric temperature decreases and the molecular bands deepen, the $(r' - $H$\alpha)$ colour of the objects increases. For carbon-rich objects the main molecular bands are due to CN or C$_2$, the deepest features of which lie longwards of 7000\AA , the long-wavelength end of the $r'$ filter. Without the TiO or VO absorption in the continuum $r'$-band, the typical ($r' - $H$\alpha$) colours of carbon stars are expected to lie below those of oxygen-rich stars by a small amount.

An intermediate chemical class of object, S-type stars can show a range of molecular absorption bands. As the C/O ratio approaches unity, bands of ZrO start to appear alongside the TiO bands, which gradually decrease. According to the spectral classification system of \citet{keen80}, ZrO bands reach a strength equal to the TiO bands at type SX/3, while ZrO is at its strongest and TiO becomes non-existent at SX/6. As the TiO bands weaken, the continuum within the $r'$-band will strengthen, but strengthening of the ZrO $\lambda6456$ feature, which partly falls within the H$\alpha$ filter, will cause the flux in this filter to decrease. The overall effect of this is for ($r' - $H$\alpha$) colour index to decrease as the ZrO bands strengthen relative to TiO. This will reach a maximum at type SX/6, where ZrO bands are at their deepest and where there are no TiO bands.

\begin{figure}
\begin{center}
\includegraphics[width=350pt, angle=270]{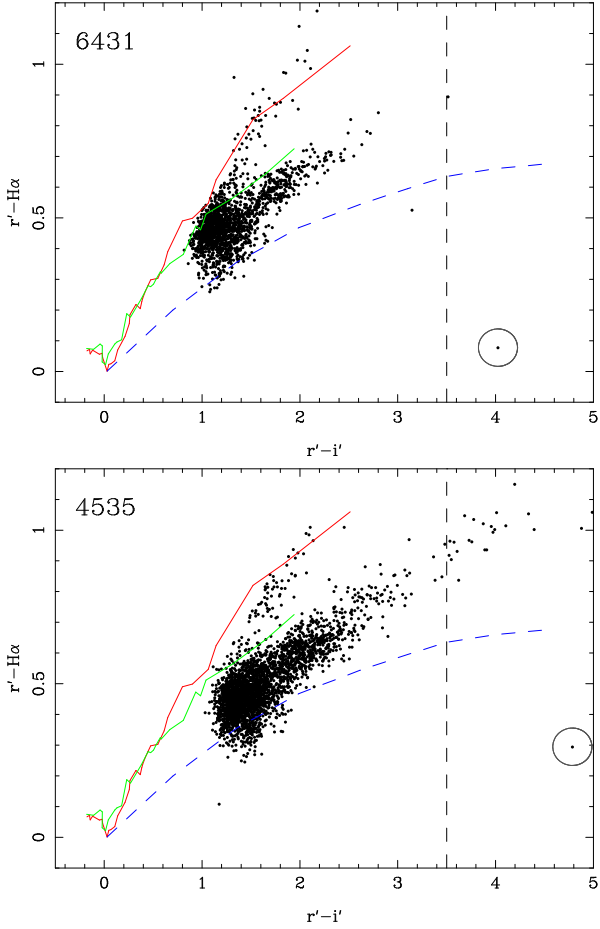}
\caption{IPHAS colour-colour diagrams for IPHAS fields 6431 and 4535, which contain the low ($r' - $H$\alpha$) index objects ERSO 76 and 80, respectively (circled objects). The sources plotted are those in the magnitude range $r' = 18-19$, a 1~magnitude interval containing the target objects. Also shown are the unreddened main-sequence (red) and giant branch (green) tracks, as well as the early-A reddening line (blue dashed line). The black vertical dashed line shows the ($r' - i') > 3.5$ limit for ERSOs.}
\label{compare_colours}
\end{center}
\end{figure}

The spectral subtype SX/6 corresponds to C/O~$\sim 0.98$ \citep{keen80}, and a large fraction of stars of this type were detected in Section~\ref{s-lowrha}. As the C/O ratio increases beyond this value the ZrO bands weaken and become invisible by SCX/8 (C/O~$\sim 1$), above which the bands of C$_2$ begin to appear. This will have the effect of increasing the ($r' - $H$\alpha$) index again, though it should not reach the levels corresponding to O-rich objects.

\subsection{A proposed selection method for chemically diverse evolved stars}

Using this information it should be possible to select these chemically diverse S-type stars using a method similar to that applied by \citet{with06} in selecting objects with an H$\alpha$ excess above the main locus in the IPHAS colour-colour diagram. The relatively reddening-indepentdent nature of the $(r' - $H$\alpha$) colour \citep{drew05} will make the separation of S-type stars in colour space particular noticeable in comparison to the effects of interstellar reddening. Figure~\ref{compare_colours} shows IPHAS colour-colour diagrams for IPHAS fields 6431 and 4535 which contain the low ($r' - $H$\alpha$) objects ERSO 76 and 80 (see Table~\ref{objects_lowrha}), both spectroscopically confirmed as S-type stars in Section~\ref{s-lowrha}.

Fields 6431 and 4535 lie in Cygnus ($l = 91.0$, $b = 1.35$) and Aquila ($l = 39.3$, $b = -1.17$) respectively, and both show clear late main-sequence and giant branches. For both fields the two S-type stars are clearly distinct at a lower ($r' - $H$\alpha$) colour than the giant branches and even below the early-A reddening line. Field 6431 also shows another potential S or C star at ($r' - $H$\alpha$, $r' - i') \sim (0.55, 3.1)$, just outside of the ERSO region. Field 4535 contains an object at ($r' - $H$\alpha$, $r' - i') \sim (0.1, 1.2)$, which, due to its bluer colours, is more likely to be a white dwarf than an AGB star \citep[white dwarfs are another class of object which can appear significantly below the main stellar loci,][]{drew05}.

A simple photometric technique for selecting reddened objects below the giant branch would be able to identify all these sources. The three objects in Table~\ref{objects_lowrha} that are not S-type stars are notably among the faintest objects in that sample and all have $r'$ magnitudes close to the magnitude limits of their observed fields and could thus be excluded from such a search.

The contributions to these sources from S~type stars and carbon stars is not yet quantifiable. The particularly low ($r' - $H$\alpha$) colours of the objects discussed in Section~\ref{s-lowrha} may have led to the selection of S-type stars over carbon stars; but the pre-identified objects discussed in Section~\ref{s-simbad} imply that carbon stars may not show such a strong photometric separation from the giant branch as do the S-type stars.

In Figure~\ref{rha_mags} the histogram of ($r' - $H$\alpha$) colour for all ERSOs provides evidence for a separate population at a lower ($r' - $H$\alpha$) colour than the main locus. The expected colours of S-type stars indicate that they could very well account for this extension to low ($r' - $H$\alpha$) index and reveals that a clear separation of populations is possible.

To assess the number of chemically diverse evolved stars that may be selected by this method, we have considered the IPHAS colour-colour diagrams for a number of fields across the galactic plane and identified those objects that lie significantly below the giant branch. We studied twenty fields in each of the regions of Aquila ($l = 40-50$), Cygnus ($l = 90-100$), Perseus ($l = 150-160$) and Taurus ($l = 180-190$, in the anticentre direction). We analysed each field at various magnitude cuts up to an appropriate magnitude limit for each field and only considered objects that were clearly separable from a well defined giant branch and had an ($r' - i'$) colour indicative of a late-type evolved star. In the four regions of Aquila, Cygnus, Perseus and Taurus, the average number of objects identified per field was 0.85, 1.05, 0.25 and 0.20 respectively.

If these count rates are integrated over the entire northern galactic plane of 7635 IPHAS fields (not including offsets since they mainly cover the same area), this would result in $\sim4700$ late-type evolved objects objects with ($r' - $H$\alpha$) colours significantly below the main giant star locus. While not all of these objects may be S~type stars (they may include highly reddened stars of an earlier spectral type, extragalactic quasi-stellar objects, carbon stars or be due to photometric errors), if even a quarter of them are S-type stars we could more than double the number of known S-type stars.

\section{Conclusions}


Over 25,000 sources were selected from the IPHAS point source catalogue based on their extremely red colours, the majority of which have no previous identifications. These extremely red stellar objects (ERSOs) are predominantly late-type giant stars. Initial follow-up spectroscopy has revealed the molecular bands indicative of cool photospheres. The distribution of these objects in the galactic plane shows strong evidence for tracing galactic extinction, indicating that their highly reddened colours are mainly due to this cause. As would be expected for late-type giant stars, a magnitude-limited study of the infrared colour excesses of these sources reveals that the majority have circumstellar material which must also be contributing to their extreme reddening.

Follow-up spectroscopy reveals that chemically diverse objects such as S-type stars will fall below O-rich evolved stars in the IPHAS colour-colour plane because of the positions of their molecular bands relative to the narrow-band H$\alpha$ filter. A method to identify these objects from the IPHAS point source catalogue is proposed and a simple estimate shows that over a thousand new S-type stars in the northern galactic plane could be discovered this way, doubling the currently known number of objects. The extension to the richer southern galactic plane using data from the forthcoming VPHAS+ survey would also be prosperous in this search.

In a further paper, we will present WHT LIRIS near-IR spectra of a large sample of ERSOs, in addition to optical spectra of ERSOs obtained using the MMT HectoSpec multi-object spectrograph. These spectra will be used to investigate the relationship between evolved star surface chemistry and photometric colours such as $(r' - $H$\alpha$).



\section{Acknowledgments}

This work is based in part on observations made with the Isaac Newton Telescope and the William Herschel Telescope, operated on the island of La Palma by the Isaac Newton Group. Observations on the WHT were obtained through an International Time Programme, awarded to the IPHAS collaboration. It also partly makes use of data products from the Two Micron All Sky Survey, the Infrared Astronomical Satellite, the Midcourse Space Experiment and the Spitzer Space Telescope, which are jointly run by the Infrared Processing and Analysis Center and the California Institute of Technology, and funded by the National Aeronautics and Space Administration and the National Science Foundation. This work also made use of observations obtained with the HectoSpec facility at the MMT Observatory, a joint facility of the Smithsonian Institution and the University of Arizona. This research has made use of the SIMBAD database, operated at CDS, Strasbourg, France. We thank the anonymous referee for their useful comments. NJW was supported by a PPARC Studentship.

\bibliographystyle{mn2e}
\bibliography{/Users/nick/Documents/Work/tex_papers/bibliography.bib}
\bsp

\label{lastpage}

\end{document}